\documentclass[acmsmall,screen]{acmart}
\pdfoutput=1
\newtheorem{remark}{Remark}
\usepackage{graphicx}
\usepackage{textcomp}
\usepackage{color}
\usepackage{xcolor}
\usepackage{epstopdf}
\usepackage{algpseudocode}
\usepackage{algorithmicx,algorithm}
\usepackage{subfigure}
\usepackage{booktabs}
\usepackage{multirow}
\usepackage{makecell}
\usepackage[T1]{fontenc}
\usepackage[utf8]{inputenc}
\usepackage{amsthm}
\usepackage{bm}
\usepackage{enumerate}
\renewcommand{\algorithmicrequire}{\textbf{Input:}}
\renewcommand{\algorithmicensure}{\textbf{Output:}}
\newtheorem{definition}{Definition} 
\newtheorem{theorem}{Theorem}
\numberwithin{equation}{section}

\renewcommand\footnotetextcopyrightpermission[1]{}
\AtBeginDocument{%
  }

\acmJournal{JACM}
\acmVolume{37}
\acmNumber{4}
\acmArticle{111}
\acmMonth{5}

\begin{document}

\title{Integrated Sensing, Communication, and Computing for Cost-effective Multimodal Federated Perception}

\author{Ning Chen}
\email{ningchen@stu.xmu.edu.cn}
\orcid{0000-0002-7364-3248}

\author{Xuwei Fan}
\email{xwfan@stu.xmu.edu.cn}

\author{Bangzhen Huang}
\email{huangbz0714@stu.xmu.edu.cn}

\author{Yifeng Zhao}
\email{zhaoyf@xmu.edu.cn}
\authornote{Corresponding author: Yifeng Zhao (zhaoyf@xmu.edu.cn).}

\author{Lianfen Huang}
\email{lfhuang@xmu.edu.cn}
\affiliation{%
  \institution{Department of Information and Communication Engineering, School of Informatics, Xiamen University}
  \city{Xiamen}
  \state{Fujian}
  \country{China}
  \postcode{361005}
}

\author{Zhipeng Cheng}
\email{chengzp_x@163.com}
\affiliation{%
  \institution{School of Future Science and Engineering, Soochow University}
  \city{Suzhou}
  \state{Jiangsu}
  \country{China}
  \postcode{215006}
}

%

%

\author{Xiaojiang Du}
\affiliation{%
  \institution{Department of Electrical and Computer Engineering, Stevens Institute of Technology}
  \city{Hoboken}
  \state{New Jersey}
  \country{USA}
  \postcode{07030}}
\email{dxj@ieee.org}

\author{Mohsen Guizani}
\affiliation{%
  \institution{Mohamed Bin Zayed University of Artificial Intelligence}
  \city{Abu Dhabi}
  \country{UAE}}
\email{mguizani@ieee.org}


\renewcommand{\shortauthors}{Ning Chen et al.}

\begin{abstract}
Federated learning (FL) is a classic paradigm of 6G edge intelligence (EI), which alleviates privacy leaks and high communication pressure caused by traditional centralized data processing in the artificial intelligence of things (AIoT). The implementation of multimodal federated perception (MFP) services involves three sub-processes, including sensing-based multimodal data generation, communication-based model transmission, and computing-based model training, ultimately relying on available underlying multi-domain physical resources such as time, frequency, and computing power. How to reasonably coordinate the multi-domain resources scheduling among sensing, communication, and computing, therefore, is crucial to the MFP networks. To address the above issues, this paper investigates service-oriented resource management with integrated sensing, communication, and computing (ISCC). With the incentive mechanism of the MFP service market, the resources management problem is redefined as a social welfare maximization problem, where the idea of ``expanding resources'' and ``reducing costs'' is used to improve learning performance gain and reduce resource costs. Experimental results demonstrate the effectiveness and robustness of the proposed resource scheduling mechanisms.
\end{abstract}





\maketitle

\section{Introduction}
The advent of the artificial intelligence of things (AIoT) era empowers the Internet of Things (IoT) with ubiquitous intelligence \cite{zhu2023pushing,you2021towards,zhang2020empowering}, while edge-AI represented by Federated learning (FL) can effectively avoid privacy leaks and reduce high communication pressure caused by data convergence in centralized learning \cite{mcmahan2017communication,chang2021survey,nguyen2021federated}. However, the intelligent connected devices in AIoT that are tasked with multimodal federated perception (MFP) services need to acquire multimodal sensing data in real-time and complete the local model training, while ensuring communication connection with the edge server to complete the model transmission, which brings a huge challenge to AIoT devices with limited time, frequency, and computing power resources \cite{chen2023joint,liu2022resource}. To this end, a cost-effective dynamic resource provisioning technology and an adaptive workload decision mechanism for MFP services are developed in this paper. 

\subsection{Motivation}
FL offers an effective solution to the training of deep learning (DL) models on data- and resource-constrained clients \cite{nguyen2021federated,gao2021federated,zhang2022federated}. However, distributed FL model training based on sensing data in AIoT still faces many challenges. In this paper, we consider three questions that also directly reveal our key motivations: i) Why can  ``\emph{MFP}'' bring benefits? ii) Why can ``\emph{extended FL}'' bring advantages? and iii) Why can ``\emph{social welfare}'' bring benefits?

The complementation and fusion of multimodal sensing technologies can expand the client's sensing domain and generate more training data samples. In recent years, DL-based computer vision (CV) technologies have achieved great success in the application of target detection in visual sensing (VS) \cite{nobis2019deep,chen2021sensing}. However, image quality is susceptible to bad weather and low-light scenes. To solve the above problems, multimodal machine learning, which combines the millimeter wave wireless sensing (WS) technology that is more robust to light and the visual sensing technology that is more sensitive to color and outline, can achieve the complementary sensing, which is a key research direction \cite{chen2021sensing,baltruvsaitis2018multimodal,nabati2019rrpn,wei2022mmwave,xiong2022unified}. In particular, integrated sensing and communication (ISAC), a candidate 6G technology, brings sensing and communication into a unified design of wireless networks, which can perform model transmission based on wireless communication (WC) while WS is in progress. Compared with single-modal visual sensing, multimodal-sensing-based MFP can expand the client's sensing domain and provide more effective data samples for FL model training \cite{liu2022vertical,de2021joint}.

Sensing for generating multimodal data samples requires the underlying physical resources, which cannot be ignored. Most of the existing research on wireless FL resource management focuses on the joint optimization between computing and communication with the assumption that the client has fixed data. However, sensing-based data generation is ignored, which is unreasonable. First, considering the clients’ need for data freshness and the adverse effects of data staleness, the data generation based on sensing and the data utilization in model training need to be closely related in executing time series, rather than severely fragmented \cite{wang2022asynchronous}. Secondly, communication and computing require physical resources, sensing for data generation also consumes clients’  physical resources, such as time and frequency \cite{zhu2023pushing,liu2022vertical}. In particular, the competitive relationship between ISAC-based wireless sensing and wireless communication over frequency resources leads to limited resource constraints, as well as the amount of data consumption is also the amount of data generated in the sensing progress, resulting in extremely complex coupling relationships and associated resource constraints among sensing, communication, and computing. Therefore, it is significant to extend the traditional FL which only considers communication and computing to extended FL, and to perform the analysis and optimization integrating sensing, communication, and computing (ISCC) by considering the coupling relationships between sensing, communication, and computing and resource constraints in the implementation of MFP.

Resource management for upper-level MFP service expects a balance of sensing, communication, and computing rather than relying on the ultimate performance of a single process. The coupling and resource constraints of MFP's sensing, communication, and computing result in local optimization of traditional single-process resource management, such as maximizing the generation rate of sensing data, maximizing communication throughput, or minimizing computing delay, is far away from the optimal resource utilization mode for learning performance. On the contrary, the global optimal model learning performance is established on the balance of sensing, communication, and computing. Meanwhile, Clients' available physical resources are heterogeneous and dynamic, which are affected by the hardware level, surrounding environment, etc. Thus, clients have different capabilities to host MFP services, as well as the distribution of data is unbalanced, which has a significant impact on the model’s learning performance \cite{zhao2018federated,liu2022joint,zhang2021adaptive}. For this, from the economic perspective, social welfare is determined by the sum of the net income of the server and working clients, which can jointly optimize multiple resource cost components and characterize the relationship between model performance and resource consumption \cite{luo2021cost}.

\subsection{Related Work}
Many existing works have been dedicated to studying resource provisioning problems in the wireless FL. An analysis model was constructed in \cite{liu2022resource} to study the relationship between the accuracy of the FL model and the resource consumption in FL-related wireless edge networks. Authors in \cite{nguyen2021toward} explored the joint resource optimization and hyper-learning rate control of the FL model. A min-max optimization problem was developed in \cite{xiao2021vehicle} to jointly optimize in-vehicle computing power, transmission power, and local model accuracy. Although considerable efforts are put into the optimization of resource management in FL, most of them focus on the joint scheduling of communication and computing resources while neglecting the resources consumed in the sensing process for data acquisition.

Among the studies of FL's resource supply strategies, the economic benefits framework of the income-expenditure perspective is an effective modeling method. Authors in \cite{luo2021cost} studied the cost-effective FL in mobile edge networks, where the time and energy costs for computing and communication are minimized. A dynamic FL economic framework for vehicle networking is proposed in \cite{saputra2021dynamic}, where authors discussed a relationship model between data quality and user environment. \cite{jiao2020toward} An auction-based market model is proposed to motivate data owners to participate in the FL. Few of the above studies consider the economic cost in the sensing process of obtaining multimodal data, which affects the fairness of transactions to some extent.

Data samples are raw materials for FL model training, and many studies have noted the impact of data quality on FL model training. Authors in \cite{liu2022joint} investigated the influence of unbalanced data distribution in hierarchical federated learning (HFL) on convergence rate and learning accuracy. A deep reinforcement learning (DRL) method is presented in \cite{zhang2021adaptive} to control training and global aggregation of local models adaptively, reducing the negative influence of non-independent identically distributed (non-IID) data on FL training performance. However, data samples are generated from environmental sensing, and clients’ ability to take on training workloads depends on their available resources. The above works ignore the relationship between data quality and the environment that the client resides in, as well as the joint impact of data quality and available physical resources on model learning performance.

To solve the problem of global optimal unreachability caused by ignoring sensing, a task-oriented resource provisioning mechanism with integrated sensing, communication, and computing has been proposed \cite{zhu2023pushing}. Authors in \cite{wen2022task} proposed an ISCC scheme for real-time inference tasks aimed to maximize inference accuracy by jointly designing sensing, quantization, and transmission under low-latency and resource constraints on the devices. There is relatively little research on the ISCC resource scheduling, and there have been some studies on the integration of two of the three entities mentioned above, including joint communication and computing, ISAC, and air computing (AirComp) \cite{zhu2023pushing}.

In this paper, compared with existing research, we take into consideration the acquisition of multimodal sensing data for MFP service in AIoT and extend the traditional FL communication round (CR) to the ISCC round (IR) that includes sensing, communication, and computing. Then the physical resources, such as time, frequency, and computing power, are decoupled from the above three processes to achieve universal resource scheduling among the sensing, communication, and computing. Furthermore, the resource management problem of MFP is modeled as maximizing social welfare in the MFP service market, achieving reasonable workload transaction decisions under the optimal resource scheduling strategy that minimizes local resource costs.


\subsection{Novelty and Contributions}

To the best of our knowledge, this paper is among the first to study the full-lifecycle universal resources management that integrates sensing, communication, and computing for the MFP service in AIoT. Major contributions are summarized as follows:

\begin{itemize}
\item {Data sample generation based on multimodal sensing.} Make full use of the complementarity of visual sensing and wireless sensing to improve the ability of multimodal data sample generation required by MFP services.
\item {Full-lifecycle resource management.} Incorporate sensing-based data sample generation into FL’s resource optimization to achieve service-oriented full-lifecycle resource management integrating sensing, communication, and computing.

\item {Cost-effective MFP service transaction mechanism.} The resource management problem is modeled to realize social welfare maximization in the MFP service market, which can be optimized from the two directions of  ``broaden sources of income and reduce expenditure''.

\item {Optimal and robust resource scheduling.} Based on convex optimization of the resource cost function, local least-cost optimal resource scheduling can be achieved, which is robust to the dynamic limited resource constraints and associated resource constraints.

\item {Simulation and analysis.} Comprehensive simulation results demonstrate the efficiency of the proposed MFP resource management mechanism integrating sensing, communication, and computing.

\end{itemize}


\section{System model and problem formulation}

In this Section, we define the universal MFP services in AIoT networks and propose the resource-efficient \underline{Z}-shaped ov\underline{er}lapped temp\underline{o}ral \underline{s}tructure (ZEROS). Finally, by developing the transaction strategies in the MFP service market, the resource management problem of MFP is modeled as the economic optimization towards maximum social welfare.

\subsection{Multimodal Federated Perception in AIoT }

\begin{figure}[htbp]
 \centerline{\includegraphics[width=5.0in]{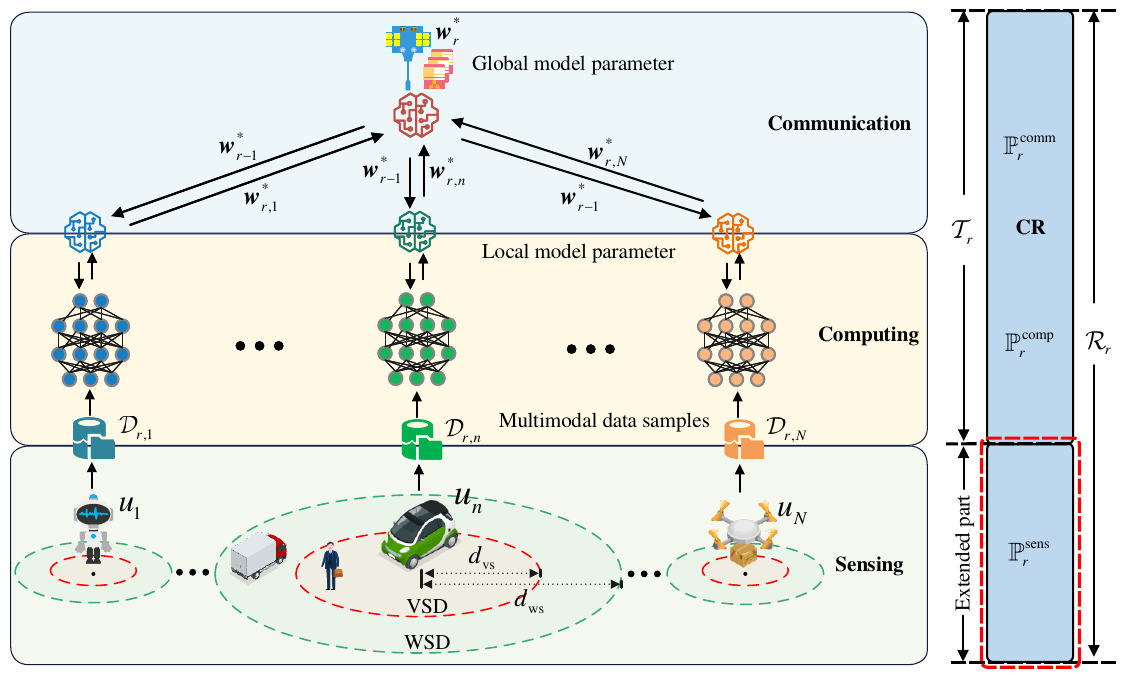}}
 \caption{MFP networks with extend FL in AIoT.}
 \label{fig1}
 \centering
\end{figure}

Considering a client-server (CS)-based MFP network, where the clients set $\mathcal{U}=\left\{ {{u}_{1}},{{u}_{2}},\ldots ,{{u}_{N}} \right\}$ is composed of $N$ AIoT devices and an edge server work as a model aggregator $m$, as shown in Fig. \ref{fig1}. Considering the differentiated resource consumption during the generation of multimodal sensing data and the impact of personalized sensing data on MFP services, we incorporate the sensing-based data generation into the optimization of FL and extend the traditional communication rounds (CRs) $\mathcal{T}=\left\{ {{\mathcal{T}}_{1}},{{\mathcal{T}}_{2}},\ldots ,{{\mathcal{T}}_{R}} \right\}$ to ISCC rounds (IRs) $\mathcal{R}=\left\{ \mathcal{R}{{}_{1}},{{\mathcal{R}}_{2}},\ldots ,{{\mathcal{R}}_{R}} \right\}$. Adopting complementary camera-based visual sensing and ISAC-based wireless sensing, we define the effective data acquisition areas as visual sensing domains (VSD) ${{S}_{\text{vs}}}$ and wireless sensing domains (WSD) ${{S}_{\text{ws}}}$ with the radius of ${{d}_{\text{vs}}}$ and ${{d}_{\text{ws}}}$, respectively, and assume that only the interested targets that appear within the sensing domain can be captured by the sensing data of corresponding modality.

\begin{figure}[t]
 \centerline{\includegraphics[width=5.0in]{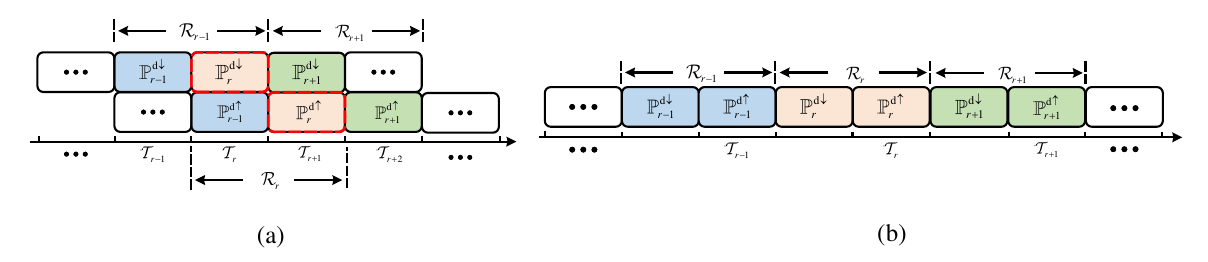}}
 \caption{Comparison between ZEROS and serial time series structure. (a) ZEROS. (b) Traditional serial time series structure. Both of them can guarantee the CSTC $\mathbb{P}_{r}^{\text{d}\downarrow }\to \mathbb{P}_{r}^{\text{d}\uparrow }$. Compared with the traditional serial time series structure, ZRTOS can achieve high reuse of time resources and improve the models’ iteration efficiency.}
 \label{fig2}
 \centering
\end{figure}

MFP service is executed in collaboration with active clients and servers. Assume the active client set in ${{\mathcal{R}}_{r}}\in \mathcal{R}$ is ${{\mathcal{U}}_{r}}=\left\{ {{u}_{n}}\in \mathcal{U}|{{\mu }_{r,n}}=1\text{ } \right\}$, where ${{\mu }_{r,n}}$ is the status indicator variable and when ${{u}_{n}}$ works in ${{\mathcal{R}}_{r}}$, ${{\mu }_{r,n}}=1$, otherwise ${{\mu }_{r,n}}=0$. Define the number of active clients in ${{\mathcal{R}}_{r}}$ is $N_{r}^{\text{act}}\triangleq \left| {{\mathcal{U}}_{r}} \right|=\sum\nolimits_{{{u}_{n}}\in \mathcal{U}}{{{\mu }_{r,n}}}\le N_{\max }^{\text{act}}$,  where $N_{\max }^{\text{act}}$ is the upper limit of the participating clients. For ${{u}_{n}}\in {{\mathcal{U}}_{r}}$ and $m$ in ${{\mathcal{R}}_{r}}\in \mathcal{R}$, considering typical FL settings, we define the universal ``\emph{MFP service}'' ${{\mathbb{P}}_{r,n}}$ in AIoT from the perspective of sensing-based multimodal data generation $\mathbb{P}_{r,n}^{{\rm{sens}}}$, communication-based model transmission $\mathbb{P}_{r,n}^{{\rm{comm}}}$ (including global model distribution $\mathbb{P}_{r,n}^{\text{comm}\downarrow }$ and local model aggregation $\mathbb{P}_{r,n}^{\text{comm}\uparrow }$), and computing-based model training $\mathbb{P}_{r,n}^{{\rm{comp}}}$ as follows \cite{mcmahan2017communication,gao2021federated}:

\begin{definition}
The active client ${{u}_{n}}\in {{\mathcal{U}}_{r}}$ generates a set of multimodal data samples ${{\mathcal{D}}_{r,n}}$ based on sensing $\mathbb{P}_{r,n}^{{\rm{sens}}}$ and puts them to the training $\mathbb{P}_{r,n}^{{\rm{comp}}}$ of the local model distributed from the server $m$ by $\mathbb{P}_{r,n}^{\text{comm}\downarrow }$ until the convergence of the local loss function $\ell ({{w}_{r,n}})$, then the obtained optimal local model $w_{r,n}^{*}$ is sent to the server by $\mathbb{P}_{r,n}^{\text{comm}\uparrow }$, and where all the active clients’ optimal local model are aggregated and output the optimal global model $w_{r}^{*}$ that minimizes the global loss function $\ell ({{w}_{r}})$. In particular, there are two compulsory serial timing constraints (CSTCs), i.e., $\mathbb{P}_{r,n}^{{\rm{comm}} \downarrow } \to\mathbb{P}_{r,n}^{{\rm{comp}}} \to \mathbb{P}_{r,n}^{{\rm{comm}} \uparrow }$ and $\mathbb{P}_{r,n}^{{\rm{sens}}} \to \mathbb{P}_{r,n}^{{\rm{comp}}}$.
\label{definition1}
\end{definition}

Ulteriorly, to solve the problem of inefficient resource reuse caused by CSTCs, we propose ZEROS, which groups the sensing $\mathbb{P}_{r,n}^{\text{sens}}$, communication $\mathbb{P}_{r,n}^{\text{comm}}$, and computing $\mathbb{P}_{r,n}^{\text{comp}}$ of MFP service ${{\mathbb{P}}_{r,n}}$ into two parts: data generation $\mathbb{P}_{r,n}^{\text{d}\downarrow }$ and data consumption $\mathbb{P}_{r,n}^{\text{d}\uparrow }$, i.e. ${{\mathbb{P}}_{r,n}}\triangleq \left\{ \mathbb{P}_{r,n}^{\text{d}\downarrow },\mathbb{P}_{r,n}^{\text{d}\uparrow } \right\}$. Among them, apparently $\mathbb{P}_{r,n}^{\text{sens}}$ is responsible for data generation, i.e. $\mathbb{P}_{r,n}^{\text{d}\downarrow }\triangleq \mathbb{P}_{r.n}^{\text{sens}}$, while  works $\mathbb{P}_{r,n}^{\text{comm}}$ and $\mathbb{P}_{r,n}^{\text{comp}}$ that constitutes a traditional CR ${{\mathcal{T}}_{r}}$ accomplishing data consumption, i.e. $\mathbb{P}_{r}^{\text{d}\uparrow }\triangleq \left\{ \mathbb{P}_{r}^{\text{comm}},\mathbb{P}_{r}^{\text{comp}} \right\}$. Obviously, there is still a CSTC $\mathbb{P}_{r}^{\text{d}\downarrow }\to \mathbb{P}_{r}^{\text{d}\uparrow }$ in the same IR. Fortunately, the foregoing circumstances will change dramatically by expanding the field of view to the adjacent IR, where the adjacent IRs are no longer executed in full-serial mode but are arranged in a Z-shaped pattern. With ZEROS, data generation $\mathbb{P}_{r}^{\text{d}\downarrow }$ of ${{\mathcal{R}}_{r}}$ and data consumption $\mathbb{P}_{r-1}^{\text{d}\uparrow }$ of ${{\mathcal{R}}_{r-1}}$ are executed in the same CR ${{\mathcal{T}}_{r}}$, which implements a cross-IR overlap similar to Tetris game, as shown in Fig. \ref{fig2}(a).

In ZEROS, each virtual IR contains two adjacent CRs, which can be expressed as:

\begin{equation}\label{e2.3}
{{\mathcal{R}}_{r}}=\left\{ {{\mathcal{T}}_{r}},{{\mathcal{T}}_{r+1}} \right\},\text{ }r=1,2,\ldots ,{{N}_{r}}
\end{equation}

We define the time consumed by the last active client to finish $\mathbb{P}_{r,n}^{\text{comm}\uparrow }$ as period $T_{\Delta }^{{{\mathcal{T}}_{r}}}$, i.e. $T_{\Delta }^{{{\mathcal{T}}_{r}}}\triangleq \max \left\{ T_{r-1,n}^{\text{d}\uparrow }|{{u}_{n}}\in {{\mathcal{U}}_{r-1}} \right\}$, where $T_{r-1,n}^{\text{d}\uparrow }$ is the time required in data consumption $\mathbb{P}_{r}^{\text{d}\uparrow }$.Given this, to maintain the continuity of CRs, the time resources $T_{r-1,n}^{\text{d}\uparrow }$ consumed in the data generation of post-IR cannot exceed the CR cycle $T_{\Delta }^{{{\mathcal{T}}_{r}}}$ identified in the pre-IR, i.e. $T_{r-1,n}^{\text{d}\uparrow }\le T_{\Delta }^{{{\mathcal{T}}_{r}}},{{u}_{n}}\in {{\mathcal{U}}_{r}}$.

According to Def. \ref{definition1}, the output optimal global model $w_{r}^{*}$ is the evaluation index of the MFP service in ${{\mathcal{R}}_{r}}\in \mathcal{R}$. The global optimization objective of MFP in ${{\mathcal{R}}_{r}}$ is defined as:

\begin{equation}\label{e2.1}
\underset{w_{r}^{*}}{\mathop{\min \ell ({{w}_{r}}):=}}\,\frac{1}{{{N}_{r}}}\underset{{{u}_{n}}\in {{\mathcal{U}}_{r}}}{\mathop{\sum }}\,{{N}_{r,n}}\ell ({{w}_{r,n}})
\end{equation}
where ${{N}_{r}}\triangleq \left| {{\mathcal{D}}_{r}} \right|$ is the size of the global set of multimodal data samples ${{\mathcal{D}}_{r}}=\sum\nolimits_{{{u}_{r,n}}\in {{\mathcal{U}}_{r}}}{{{\mathcal{D}}_{r,n}}}$, ${N_{r,n}}$ represents the number of samples contributed by client ${{u}_{n}}$.

Based on the above analysis, the optimal global model depends on the convergence of the active clients' local model. The local optimization objective of active client ${{u}_{n}}$ in IR ${{\mathcal{R}}_{r}}\in \mathcal{R}$ can be defined as:

\begin{equation}\label{e2.2}
\underset{w_{r,n}^{*}}{\mathop{\min \ell ({{w}_{r,n}})}}\,:=\frac{1}{{{N}_{r,n}}}\sum\limits_{k=1}^{{{N}_{r,n}}}{\ell \left( {{w}_{r,n}},{{\xi }_{r,n,k}} \right)}
\end{equation}
where ${{\xi }_{r,n,k}}\triangleq \left( {{\mathbf{x}}_{r,n,k}},{{y}_{r,n,k}} \right)$ is the multimodal data sample in ${{\mathcal{R}}_{r}}$, ${{\mathbf{x}}_{r,n,k}}\triangleq \left( \mathbf{x}_{r,n,k}^{\text{vs}},\mathbf{x}_{r,n,k}^{\text{ws}} \right),\text{ }1\le k\le {{N}_{r,n}}$ is the $k\text{-th}$ visual-wireless-fused feature (hereinafter referred to as fused feature), where $\mathbf{x}_{r,n,k}^{\text{vs}}$ and $\mathbf{x}_{r,n,k}^{\text{ws}}$ is the visual feature and wireless feature, respectively. ${{y}_{r,n,k}}\in \mathcal{Y},\text{ }1\le k\le {{N}_{r,n}}$ is the $k\text{-th}$ multimodal data sample label, where the alphabet $\mathcal{Y}=\left\{ {{a}_{1}},{{a}_{2}},\cdots ,{{a}_{|\mathcal{Y}|}} \right\}$ represents the label space, such as the set of classes in the target classification or recognition task \cite{liu2022joint,nguyen2021toward,wang2020optimizing,cover1999elements}.


%

\subsection{MFP Service Market with Integrated Sensing, Communication, and Computing}

\begin{figure}[t]
 \centerline{\includegraphics[width=5.5in]{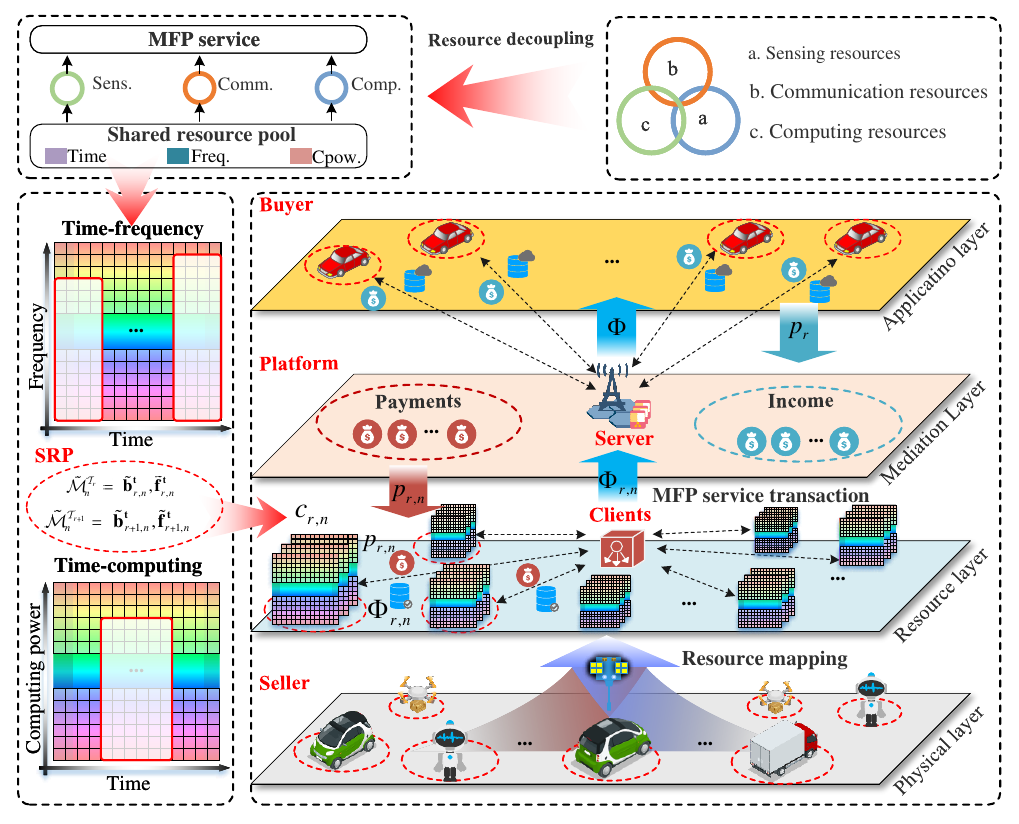}}
 \caption{Economic model of the MFP service market based on SRP.}
 \centering
 \label{fig3}
\end{figure}

Defines that the MFP service market consists of sellers, platforms, and buyers. Among them, working clients play the role of the seller, which consumes local time, frequency, and, computing power resources to complete the data generation (i.e., sensing) and data consumption (including communication and computing) of MFP services. A buyer is an upper AI application in the application layer that accepts the output optimal model of MFP and pays for the model learning performance gain obtained. The server  is the platform that acts as the intermediary and serves connecting buyers and sellers, receiving buyers’ payment for model performance gains and paying sellers the cost of workload measured in data volume. The economic model of the MFP service market is shown in Fig. \ref{fig3}.

To find the MFP service-oriented optimal resource scheduling strategies, we propose a the shared resource pool (SRP) model that decouples the multi-domain physical resources from the processes composed by sensing, communication, and computing, as shown in Fig. \ref{fig3}.Notably, SRPs are shared by adjacent IRs and are not exclusive to either party of the sensing, communication, and computing, but they can use the resources with equal status. Furthermore, since both frequency and computing power resources have time attributes, we integrate one-dimensional time resources into the above two to get two-dimensional time-frequency resource pools and time-computation resource pools.

Specifically, the set of available one-dimensional resource vectors for client ${{u}_{n}}\in \mathcal{U}$ in CR ${{\mathcal{T}}_{r}}\in \mathcal{T}$ can be defined as:

\begin{equation}\label{e3.1}
\tilde{\mathcal{M}}_{r,n}^{\text{I}}=\left\{ {{{\mathbf{\tilde{t}}}}_{r,n}},{{{\mathbf{\tilde{b}}}}_{r,n}},{{{\mathbf{\tilde{f}}}}_{r,n}} \right\}
\end{equation}
where ${{\mathbf{\tilde{t}}}_{r,n}}$, ${{\mathbf{\tilde{b}}}_{r,n}}$ and ${{\mathbf{\tilde{f}}}_{r,n}}$ are the time resource vector, the frequency resource vector and the computing power resource vector, respectively, whose one-dimensional discrete quantization scales are $\tilde{T}_{n}^{{{\mathcal{T}}_{r}}}\triangleq \left| {{{\mathbf{\tilde{t}}}}_{r,n}} \right|$ (s), $\tilde{B}_{n}^{{{\mathcal{T}}_{r}}}\triangleq \left| {{{\mathbf{\tilde{b}}}}_{r,n}} \right|$ (Hz), $\tilde{F}_{n}^{{{\mathcal{T}}_{r}}}\triangleq \left| {{{\mathbf{\tilde{f}}}}_{r,n}} \right|$ (CPU cycle), respectively.

Due to the time attributes of frequency and computing, we further construct a two-dimensional resource matrix, i.e. time-frequency matrix $\mathbf{\tilde{b}}_{r,n}^{\mathbf{t}}\triangleq {{\mathbf{O}}^{\tilde{B}_{n}^{{{\mathcal{T}}_{r}}}\times \tilde{T}_{n}^{{{\mathcal{T}}_{r}}}}}$ and time-computing matrix $\mathbf{\tilde{f}}_{r,n}^{\mathbf{t}}\triangleq {{\mathbf{O}}^{\tilde{F}_{n}^{{{\mathcal{T}}_{r}}}\times \tilde{T}_{n}^{{{\mathcal{T}}_{r}}}}}$. Thus, the SRP of the active client ${{u}_{n}}\in \mathcal{U}$ in CR ${{\mathcal{T}}_{r}}\in \mathcal{T}$ can be described as:

\begin{equation}\label{e3.2}
\tilde{\mathcal{M}}_{n}^{{{\mathcal{T}}_{r}}}=\left\{ \mathbf{\tilde{b}}_{r,n}^{\mathbf{t}},\mathbf{\tilde{f}}_{r,n}^{\mathbf{t}} \right\}
\end{equation}
where $\mathbf{\tilde{b}}_{r,n}^{\mathbf{t}}$ and $\mathbf{\tilde{f}}_{r,n}^{\mathbf{t}}$ in the initial SRPs are all-zero matrices, and it will become to 1 when the resource represented by an element in them is consumed.

In the MFP service market, the revenue of the middleman server comes from the payments made by upper AI applications for MFP services, which focus only on the learning performance gains of the resulting models. Consequently, the income ${{p}_{r}}$ in ${{\mathcal{R}}_{r}}$, i.e. the payment from applications, can be calculated as:

\begin{equation}\label{e3.3}
{{p}_{r}}={{\Delta }_{\text{ }\!\!\theta\!\!\text{ }}}{{\Phi }_{r}}
\end{equation}
where ${{\Delta }_{\text{ }\!\!\theta\!\!\text{ }}}$ is the unit price of the model learning performance gain and ${{\Phi }_{r}}$ is the global model learning gain in ${{\mathcal{R}}_{r}}$, which can be expressed as:

\begin{equation}\label{e3.4}
{{\Phi }_{r}}\triangleq \sum\nolimits_{{{u}_{n}}\in \mathcal{U}}{{{\Phi }_{r,n}}}=\sum\nolimits_{{{u}_{n}}\in \mathcal{U}}{\left( {{\lambda }_{\text{s-p}}}\mathcal{Q}\left( r,n \right)N_{r,n}\right)}
\end{equation}
where ${{\Phi }_{r,n}}$ is the learning performance gain contributed by client ${u_n}$ in IR ${{\mathcal{R}}_{r}}$, \cite{cheng2023learning,jiao2020toward}, ${\lambda _{{\rm{s - p}}}}$ is the learning performance factors of IID data samples,  and  $Q\left( {r,n} \right)$ is the quality of data (QoD), which is determined by the difference between client' local distribution and global distribution of multimodal data samples and can be expressed as ${{\mathsf{\mathcal{Q}}}_{r,n}}:=1-\mathbb{E}\left[ \sum\limits_{c\in \mathcal{Y}}{\left\| {{p}_{r,n,c}}-{{{\bar{p}}}_{r,c}} \right\|} \right]$ \cite{wang2022asynchronous,zhao2018federated,liu2022joint,jiao2020toward,jiao2018profit}.


The payment ${{p}_{r,n}}$ that the server $\text{m}$ made to the active clients ${{u}_{n}}$ is only related to the MFP workload measured by the number of multimodal data samples, i.e.:

\begin{equation}\label{e3.5}
{{p}_{r,n}}={{\Delta }_{\text{s}}}{{N}_{r,n}}
\end{equation} 
where ${{\Delta }_{\text{s}}}$ is the unit price of multimodal data samples.

The working client executes MFP services consuming resources in the local SRP \cite{luo2021cost}. Define the set of unit prices of time, frequency, and computing power resource as:

\begin{equation}\label{e3.6}
\Delta =\left\{ {{\Delta }_{\text{t}}},{{\Delta }_{\text{b}}},{{\Delta }_{\text{f}}} \right\}
\end{equation} 


%

Thus, the total resource cost ${{c}_{r,n}}$ of the client ${{u}_{n}}$ in IR ${{\mathcal{R}}_{r}}$ can be calculated as:

\begin{equation}\label{e3.7}
{{c}_{r,n}}=c_{r,n}^{\text{t}}+c_{r,n}^{\text{b}}+c_{r,n}^{\text{f}}
\end{equation} 
where $c_{r,n}^{\text{t}}\triangleq {{T}_{r,n}}{{\Delta }_{\text{t}}}$, $c_{r,n}^{\text{b}}\triangleq {{B}_{r,n}}{{\Delta }_{\text{b}}}$, and $c_{r,n}^{\text{f}}\triangleq {{F}_{r,n}}{{\Delta }_{\text{f}}}$ are the costs of time, frequency, and computing resources, respectively. The method for determining the amount of resources consumed in SRP is shown in Def. \ref{definition2}.

\begin{definition}
The resource consumption of the type $X\in \left\{ T,B,F \right\}$ of client ${{u}_{n}}$ in IR ${{\mathcal{R}}_{r}}$ is defined as ${{X}_{r,n}}\triangleq X_{r,n}^{{{\mathcal{T}}_{r}}}+X_{r,n}^{{{\mathcal{T}}_{r+1}}}=\bigcup {{\mathbb{P}}_{r}}\left( \left| {{\mathbf{r}}_{X}} \right| \right)$, $\mathbf{r}\in \tilde{\mathcal{M}}_{n}^{{{\mathcal{T}}_{r}}}\bigcup \tilde{\mathcal{M}}_{n}^{{{\mathcal{T}}_{r+1}}}$, which is the sum of the number of elements $X_{r,n}^{{{\mathcal{T}}_{r}}}$ and $X_{r,n}^{{{\mathcal{T}}_{r+1}}}$ in the $X$ direction of SRP $\tilde{\mathcal{M}}_{n}^{{{\mathcal{T}}_{r}}}=\left\{ \mathbf{\tilde{b}}_{r,n}^{\mathbf{t}},\mathbf{\tilde{f}}_{r,n}^{\mathbf{t}} \right\}$ and $\tilde{\mathcal{M}}_{n}^{{{\mathcal{T}}_{r+1}}}=\left\{ \mathbf{\tilde{b}}_{r+1,n}^{\mathbf{t}},\mathbf{\tilde{f}}_{r+1,n}^{\mathbf{t}} \right\}$ that are first occupied during data generation $\mathbb{P}_{r,n}^{\text{d}\downarrow }$ (i.e. sensing $\mathbb{P}_{r,n}^{\text{sens}}$) and data consumption $\mathbb{P}_{r,n}^{\text{d}\uparrow }$ (including communication $\mathbb{P}_{r,n}^{\text{comm}}$ and computing $\mathbb{P}_{r,n}^{\text{comp}}$) during the execution of MFP service ${{\mathbb{P}}_{r}}$, where ${{\mathbb{P}}_{r}}\left( \left| {{\mathbf{r}}_{X}} \right| \right)$ represents the number of elements occupied by MFP service ${{\mathbb{P}}_{r}}$ in the $X\text{-direction}$ resource type in SRP $\mathbf{r}$.
\label{definition2}
\end{definition}


Based on the discussion above, the net profit $R_{r}^{\text{m}}$ of the server $\text{m}$ in ${{\mathcal{R}}_{r}}$ is:

\begin{equation}\label{e3.8}
R_{r}^{\text{m}}={{p}_{r}}-\sum\limits_{{{u}_{n}}\in \mathcal{U}}{{{p}_{r,n}}}
\end{equation}

Similarly, the net profit ${{R}_{r,n}}$ of the client ${{u}_{n}}$ is:

\begin{equation}\label{e3.9}
{{R}_{r,n}}={{p}_{r,n}}-{{c}_{r,n}}
\end{equation}

%
%
%
%
%
%
%
%

\subsection{Problem Formulation}

Based on the above definition and analysis, we define the optimization goal of MFP resource management as determining the optimal strategy ${{\pi }^{*}}$  that maximizes the social welfare of MFP service transactions while meeting the learning performance gain requirements of MFP model \cite{xiong2022unified,wang2020optimizing,jiao2020toward}:

\begin{equation}\label{e3.12}
\mathbf{P1:}\underset{{{\pi }^{*}}}{\mathop{\arg \max }}\,R\triangleq \sum\limits_{{{\mathcal{R}}_{r}}\in \mathcal{R}}{{{R}_{r}}}=\sum\limits_{{{\mathcal{R}}_{r}}\in \mathcal{R}}{\left( \alpha \cdot R_{r}^{\text{m}}+\beta \sum\limits_{{{u}_{n}}\in {{\mathcal{U}}_{r}}}{{{R}_{r,n}}} \right)}
\end{equation} 
s.t.

\begin{equation}\label{e2.6}
R_{r}^{\text{m}}\ge 0,\text{ }{{R}_{r,n}}\ge 0
\end{equation}

\begin{equation}\label{e2.5}
{{\theta }_{0}}\le \Phi \triangleq \sum\limits_{{{\mathcal{R}}_{r}}\in \mathcal{R}}{\sum\limits_{{{u}_{n}}\in \mathcal{U}}{{{\Phi }_{r,n}}}}<{{\theta }_{0}}+{{\Delta }_{\theta }}
\end{equation}

\begin{equation}\label{e2.7}
T_{r-1,n}^{\text{d}\uparrow }\triangleq T_{r-1,n}^{\text{comm}\downarrow }+T_{r-1,n}^{\text{comp}}+T_{r-1,n}^{\text{comm}\uparrow }\le \min \left\{ \tilde{T}_{n}^{{{\mathcal{T}}_{r}}},T_{\Delta }^{{{\mathcal{T}}_{r}}} \right\}
\end{equation}

\begin{equation}\label{e2.8}
T_{r,n}^{\text{d}\downarrow }\triangleq T_{r,n}^{\text{sens}}\le \min \left\{ \tilde{T}_{n}^{{{\mathcal{T}}_{r}}},T_{\Delta }^{{{\mathcal{T}}_{r}}} \right\}
\end{equation}


\begin{equation}\label{e2.10}
B_{r,n}^{\mathcal{T}}\triangleq B_{r,n}^{\text{sens}}+B_{r-1,n}^{\text{comm}\downarrow }+B_{r-1,n}^{\text{comm}\uparrow }
\end{equation}

\begin{equation}\label{e2.11}
\left( B_{r,n}^{\text{sens}},B_{r-1,n}^{\text{comm}\downarrow },B_{r-1,n}^{\text{comm}\uparrow } \right)|T\le \tilde{B}_{n}^{{{\mathcal{T}}_{r}}}
\end{equation}

\begin{equation}\label{e2.12}
F_{r-1,n}^{\text{comp}}\le \tilde{F}_{n}^{{{\mathcal{T}}_{r}}}
\end{equation}

\begin{equation}\label{e2.13}
{{N}_{r,n}}=\sum\limits_{c\in \mathcal{Y}}{{{N}_{r,n,c}}}\triangleq \underbrace{Output\left( \mathbb{P}_{r,n}^{\text{sens}}\left( T_{r,n}^{\text{sens}},B_{r,n}^{\text{sens}} \right) \right)}_{\text{Perspective of Sensing}}\Leftrightarrow \underbrace{Input\left( \mathbb{P}_{r,n}^{\text{comp}}\left( T_{r,n}^{\text{comp}},F_{r,n}^{\text{comp}} \right) \right)}_{\text{Perspective of Computing}}
\end{equation}
where ${{R}_{r}}\triangleq \alpha \cdot R_{r}^{\text{m}}+\beta \sum\nolimits_{{{u}_{n}}\in {{\mathcal{U}}_{r}}}{{{R}_{r,n}}}$ is the social welfare of IR ${{\mathcal{R}}_{r}}$, where $\alpha $ is the benefit weight of the server while $\beta $ is the benefit weight of the client \cite{luo2021cost}. $\tilde{\mathcal{M}}_{n}^{{{\mathcal{T}}_{r}}}=\left\{ \tilde{T}_{n}^{{{\mathcal{T}}_{r}}},\tilde{B}_{n}^{{{\mathcal{T}}_{r}}},\tilde{F}_{n}^{{{\mathcal{T}}_{r}}} \right\}$ are the local available resources, which can be determined by the client work scenario, such as the frequency resources of AIoT devices for narrowband communication are scarce, while low-cost AIoT devices are usually not equipped with a large amount of computing resources. Correspondingly, $\mathcal{M}_{n}^{{{\mathcal{T}}_{r}}}=\left\{ T_{n}^{{{\mathcal{T}}_{r}}},B_{n}^{{{\mathcal{T}}_{r}}},F_{n}^{{{\mathcal{T}}_{r}}} \right\}$ are the resource consumption, which includes both resource consumption about $\mathbb{P}_{r-1,n}^{\text{d}\uparrow }$ in ${{\mathcal{R}}_{r-1}}$ and resource consumption about $\mathbb{P}_{r,n}^{\text{d}\downarrow }$ in ${{\mathcal{R}}_{r}}$. Constraints \ref{e2.6} specifies the economic attributes of individual rationality in the transaction activities of the MFP service market; Constraints \ref{e2.5} indicate that learning performance gain of the MFP model needs to be located in the feasible interval $\left[ {{\theta }_{0}},{{\theta }_{0}}+{{\Delta }_{\theta }} \right)$, where ${{\Delta }_{\theta }}$ is the discrete granularity of learning performance gain; Constraints \ref{e2.7} represent the time resources consumed by data consumption $\mathbb{P}_{r-1,n}^{\text{d}\uparrow }$ of ${{\mathcal{R}}_{r-1}}$ cannot exceed the constraints of available time resources and CR cycles; Similarly, constraints \ref{e2.8} represent constraints on time resources consumed by data generation. Constraints \ref{e2.10} define the composition of frequency resource consumption; Constraints \ref{e2.11} are caused by the ISAC spectrum sharing, which represents the sum of the frequency resources consumed by the time resource multiplexing portion cannot exceed the available frequency resources. Constraints \ref{e2.12} define that the consumption of computing power resources cannot exceed the available resources. Constraints \ref{e2.13} expounds the equality constraints arising from the same workload of sensing and computing in the same IR.

%
%

The problem \emph{P1} can be optimized from the two directions of  "broaden sources of income and reduce expenditure". First, "broaden sources of income" is to increase the gross profit of the market, which is the payment of upper applications for model learning performance gains, i.e., to maximize the MFP learning performance gain aggregated from all working clients. Second, "reduce expenditure" is to achieve the minimum resource cost for a given workload. In the following section, to simplify the analysis, from the perspective of a single IR that contains two adjacent CRs, we prove that the local resource cost minimization in the MFP service market is a convex optimization problem, and then give a robust optimal resource scheduling strategy under dynamic resource constraints.

\section{Multiplier-based Optimal Universal Resource Scheduling}

This section demonstrates that exploring optimal local resource scheduling in the MFP service market can be modeled as a convex optimization problem about minimizing resource costs, whether with resource constraints or without. Furthermore, by the Lagrange-multiplier method, the minimum problem with inequality constraints is transformed into an unconstrained optimization problem for Lagrange functions that satisfy the Karush-Kuhn-Tucker (KKT) Conditions. The multiplier-based optimal universal resource scheduling (M-OURS) algorithm is proposed, which offers a robust optimal local resource scheduling strategy with dynamic resource constraints.

\subsection{Unconstrained Universal Resource Scheduling}

The local resource scheduling problem of the client ${{u}_{n}}$ in IR ${{\mathcal{R}}_{r}}\in \mathcal{R}$ that minimizes resource costs for MFP service transactions with the workload ${{N}_{r,n}}$, can be modeled as:

\begin{equation}\label{e4.1}
\mathbf{P2:}\underset{\varpi _{r,n}^{*}}{\mathop{\text{arg min:}}}\,{{c}_{r,n}}\left( {{N}_{r,n}} \right)=c_{r,n}^{\text{t}}\left( {{N}_{r,n}} \right)+c_{r,n}^{\text{b}}\left( {{N}_{r,n}} \right)+c_{r,n}^{\text{f}}\left( {{N}_{r,n}} \right)
\end{equation} 
where ${{\varpi }_{r,n}}$ is the local resource scheduling policy and $\varpi _{r,n}^{*}$ is the optimal resource scheduling policy with minimum resource costs.

Based on the temporal structure ZEROS proposed in this paper, if the client ${{u}_{n}}$ assumes the MFP workload in IR ${{\mathcal{R}}_{r}}$, data generation $\mathbb{P}_{r,n}^{\text{d}\downarrow }$ (i.e., sensing $\mathbb{P}_{r,n}^{\text{sens}}$) needs to be performed in CR ${{\mathcal{T}}_{r}}$, and data consumption $\mathbb{P}_{r,n}^{\text{d}\uparrow }$ (including global model distribution $\mathbb{P}_{r,n}^{\text{comm}\downarrow }$, computing $\mathbb{P}_{r,n}^{\text{comp}}$, and local model upload $\mathbb{P}_{r,n}^{\text{comm}\uparrow }$) needs to be performed in CR ${{\mathcal{T}}_{r+1}}$. Therefore, with the given MFP workload ${{N}_{r,n}}$, the resource cost of the client ${{u}_{n}}$ in ${{\mathcal{R}}_{r}}$ is the sum of its resource cost in ${{\mathcal{T}}_{r}}$ and ${{\mathcal{T}}_{r+1}}$. Since SRP $\tilde{\mathcal{M}}_{n}^{{{\mathcal{T}}_{r}}}=\left\{ \mathbf{\tilde{b}}_{r,n}^{\mathbf{t}},\mathbf{\tilde{f}}_{r,n}^{\mathbf{t}} \right\}$ and $\tilde{\mathcal{M}}_{n}^{{{\mathcal{T}}_{r+1}}}=\left\{ \mathbf{\tilde{b}}_{r+1,n}^{\mathbf{t}},\mathbf{\tilde{f}}_{r+1,n}^{\mathbf{t}} \right\}$ of ${{\mathcal{T}}_{r}}$ and ${{\mathcal{T}}_{r+1}}$  are independent, the problem of minimum resource cost can be equivalent to the minimum resource cost $c_{r,n}^{\text{d}\downarrow }$ of data generation $\mathbb{P}_{r,n}^{\text{d}\downarrow }$ and the minimum resource cost $c_{r,n}^{\text{d}\uparrow }$ of data consumption $\mathbb{P}_{r,n}^{\text{d}\uparrow }$, respectively. As a result, \emph{P3} is equivalent as:

\begin{equation}\label{e4.2}
\underset{\varpi _{r,n}^{*}}{\mathop{\text{arg min:}}}\,{{c}_{r,n}}\left( {{N}_{r,n}} \right)\Leftrightarrow \underset{\varpi _{r,n,r}^{*}}{\mathop{\text{arg min:}}}\,c_{r,n}^{\text{d}\downarrow }\left( {{N}_{r,n}} \right)+\underset{\varpi _{r,n,r+1}^{*}}{\mathop{\text{arg min:}}}\,c_{r,n}^{\text{d}\uparrow }\left( {{N}_{r,n}} \right)
\end{equation}
where ${{\varpi }_{r,n,r}}$ and ${{\varpi }_{r,n,r+1}}$ are the local resource scheduling strategies of the client ${{u}_{n}}$ in CR ${{\mathcal{T}}_{r}}$ and ${{\mathcal{T}}_{r+1}}$, respectively, which satisfies ${{\varpi }_{r,n}}=\left\{ {{\varpi }_{r,n,r}},{{\varpi }_{r,n,r+1}} \right\}$.

With the condition that the client has no constraints, that is, the available resources in its SRP $\tilde{\mathcal{M}}_{n}^{{{\mathcal{T}}_{r}}}=\left\{ \mathbf{\tilde{b}}_{r,n}^{\mathbf{t}},\mathbf{\tilde{f}}_{r,n}^{\mathbf{t}} \right\}$ and $\tilde{\mathcal{M}}_{n}^{{{\mathcal{T}}_{r+1}}}=\left\{ \mathbf{\tilde{b}}_{r+1,n}^{\mathbf{t}},\mathbf{\tilde{f}}_{r+1,n}^{\mathbf{t}} \right\}$ are infinite, and without considering the resource relationship constraints \ref{e2.13} of the adjacent CRs, we derive Thm. \ref{theorem1} as follows:

\begin{theorem}
The unconstrained resource cost minimization problem for clients ${{u}_{n}}$ in IR ${{\mathcal{R}}_{r}}\in \mathcal{R}$ is a convex optimization problem. For any workload ${{N}_{r,n}}$, there is only one optimal resource scheduling strategy $\varpi _{r,n}^{*}\left( {{N}_{r,n}} \right)$ that minimizes the cost of local resources.
\label{theorem1}
\end{theorem}

The proof of Thm. \ref{theorem1} can be found in Appendix \ref{A}.

\subsection{Constrained Universal Resource Scheduling}

\begin{algorithm}[htbp]  
	\renewcommand{\algorithmicrequire}{\textbf{Input:}}
	\renewcommand{\algorithmicensure}{\textbf{Output:}}
	\caption{The proposed M-OURS algorithm.}  
	\label{algorithm1}
	\begin{algorithmic}[1] 
		\Require The set of workload ${{N}_{r}}=\left\{ {{N}_{r,n}},n=1,2,\ldots ,N \right\}$, the set of SRPs ${{\tilde{\mathcal{M}}}_{r}}=\left\{ \left\{ ~\tilde{\mathcal{M}}_{n}^{{{\mathcal{T}}_{r}}}=\left\{ \mathbf{\tilde{b}}_{r,n}^{\mathbf{t}},\mathbf{\tilde{f}}_{r,n}^{\mathbf{t}} \right\},\tilde{\mathcal{M}}_{n}^{{{\mathcal{T}}_{r+1}}}=\left\{ \mathbf{\tilde{b}}_{r+1,n}^{\mathbf{t}},\mathbf{\tilde{f}}_{r+1,n}^{\mathbf{t}} \right\} \right\},n=1,2,\ldots ,N \right\}$, and the set of resource scheduling strategies in pre-IR ${{\varpi }_{r-1}}=\left\{ \varpi _{r-1,n}^{*},n=1,2,\ldots ,N \right\}$.
		\Ensure  Optimal resource scheduling strategies $\varpi _{r}^{*}=\left\{ \varpi _{r,n}^{*}\left( {{N}_{r,n}} \right),n=1,2,\ldots ,N \right\}$.

		\For{$n$ in range $N$} 
			\If{${{\mu }_{r,n}}=1$}
				\State Evaluate resources and workload:
				\State ${{N}_{r,n}}\rightleftharpoons \left\{ ~\tilde{\mathcal{M}}_{n}^{{{\mathcal{T}}_{r}}}=\left\{ \mathbf{\tilde{b}}_{r,n}^{\mathbf{t}},\mathbf{\tilde{f}}_{r,n}^{\mathbf{t}} \right\},\tilde{\mathcal{M}}_{n}^{{{\mathcal{T}}_{r+1}}}=\left\{ \mathbf{\tilde{b}}_{r+1,n}^{\mathbf{t}},\mathbf{\tilde{f}}_{r+1,n}^{\mathbf{t}} \right\} \right\}|{{\varpi }_{r-1,n}}$
				\State Calculate the status attributes ${{S}_{r,n}}$: ${{S}_{r,n}}=\left\{ {{a}_{r,n}},{{b}_{r,n}},{{c}_{r,n}},{{d}_{r,n}},{{e}_{r,n}} \right\}$
				\State Calculate the maximum transaction volume $N_{r,n}^{\max }$:
				\State $N_{r,n}^{\max }=\min \left\{ N_{r,n}^{\max }\left( \mathbb{P}_{r,n}^{\text{d}\downarrow } \right),N_{r,n}^{\max }\left( \mathbb{P}_{r,n}^{\text{d}\uparrow } \right) \right\}$
				\If{${{N}_{r,n}}\le N_{r,n}^{\max }$}
					\If{${{N}_{r,n}}\in \left\{ 0,1,2,\ldots ,N_{r,n}^{\text{unc}} \right\}$}
						\State Attributable to Thm. \ref{theorem1}
						\State $\min \left( {{c}_{r,n}}|{{N}_{r,n}} \right)\triangleq \min \left( c_{r,n}^{\text{d}\downarrow }|{{N}_{r,n}} \right)+\min \left( c_{r,n}^{\text{d}\uparrow }|{{N}_{r,n}} \right)$
						\State Calculate $\varpi _{r,n,r}^*\left( {{N_{r,n}}} \right): \Leftarrow \mathop {{\rm{arg min:}}}\limits_{\varpi _{r,n,r}^*} c_{r,n}^{{\rm{d}} \downarrow }$, $\varpi _{r,n,r + 1}^*\left( {{N_{r,n}}} \right): \Leftarrow \mathop {{\rm{arg min:}}}\limits_{\varpi _{r,n,r + 1}^*} c_{r,n}^{{\rm{d}} \uparrow }$
						\State \underline{Solution 1}: $\varpi _{r,n}^*\left( {{N_{r,n}}} \right) \Leftarrow \left\{ {\varpi _{r,n,r}^*\left( {{N_{r,n}}} \right),\varpi _{r,n,r + 1}^*\left( {{N_{r,n}}} \right)} \right\}$
					\Else { ${{N}_{r,n}}\in \left\{ N_{r,n}^{\text{unc}}+1,N_{r,n}^{\text{unc}}+2,\ldots ,N_{r,n}^{\max } \right\}$}
						\State Attributable to Thm. \ref{theorem2}
						\State Construct Lagrangian functions: $\mathcal{L}(\mathbf{x},\mathbf{y},\mathbf{z},\mathbf{\mu },\mathbf{\lambda })$
						\State Calculate $\mathcal{L}$ with KKT conditions 
                                \State \underline{Solution 2}: $\varpi _{r,n}^{*}\left( {{N}_{r,n}} \right)\Leftarrow \mathcal{L}(\mathbf{x},\mathbf{y},\mathbf{z},\mathbf{\mu },\mathbf{\lambda })|\text{KKT conditions}$
					\EndIf
				\Else { ${{N}_{r,n}}>N_{r,n}^{\max }$}
				\State No available resource scheduling strategy
                     \State \underline{Solution 3}: $\varpi _{r,n}^{*}\left( {{N}_{r,n}} \right)=\varnothing $
				\EndIf
			\Else { ${{\mu }_{r,n}}=0$}
				\State Not participating in transactions
                     \State \underline{Solution 4}: $\varpi _{r,n}^{*}\left( {{N}_{r,n}} \right)=\varnothing $
			\EndIf
		\State \textbf{Output } $\varpi _{r,n}^{*}\left( {{N}_{r,n}} \right)$
		\EndFor 
	\end{algorithmic}
\end{algorithm}

Thm. \ref{theorem1} illustrates that the unconstrained minimizing resource cost problem has only one solution and can be solved by convex optimization. However, in real-world scenarios, SRPs $\tilde{\mathcal{M}}_{n}^{{{\mathcal{T}}_{r}}}=\left\{ \mathbf{\tilde{b}}_{r,n}^{\mathbf{t}},\mathbf{\tilde{f}}_{r,n}^{\mathbf{t}} \right\}$ and $\tilde{\mathcal{M}}_{n}^{{{\mathcal{T}}_{r+1}}}=\left\{ \mathbf{\tilde{b}}_{r+1,n}^{\mathbf{t}},\mathbf{\tilde{f}}_{r+1,n}^{\mathbf{t}} \right\}$ are limited, and clients need to perform local resource scheduling with limited resources. Furthermore, there are also resource constraints \ref{e2.13} on equations for adjacent CRs. Therefore, based on \emph{P3}, the local resource cost minimization problem under resource constraints can be modeled as:

\begin{equation}\label{e4.3}
\mathbf{P3:}\underset{\varpi _{r,n}^{*}}{\mathop{\text{arg min:}}}\,{{f}_{r,n}}\left( \mathbf{x},\mathbf{y},\mathbf{z} \right)={{\Delta }_{\text{t}}}\sum\limits_{{{x}_{i}}\in \mathbf{x}}{{{x}_{i}}}+{{\Delta }_{\text{b}}}\sum\limits_{{{y}_{i}}\in \mathbf{y}}{{{y}_{i}}}+{{\Delta }_{\text{f}}}\sum\limits_{{{z}_{i}}\in \mathbf{z}}{{{z}_{i}}}
\end{equation}
S.T.

\begin{equation}\label{e4.4}
\mathbf{h}\left( \mathbf{x},\mathbf{y},\mathbf{z} \right)\triangleq \left\{ \underbrace{{{h}_{i}}\left( \mathbf{x},\mathbf{y},\mathbf{z} \right)=0}_{\text{Equality constraint}}\text{ }\Leftarrow \text{Constraint}(2.13),\text{ }i=1,2,\ldots ,H \right\}
\end{equation}

\begin{equation}\label{e4.5}
\mathbf{g}\left( \mathbf{x},\mathbf{y},\mathbf{z} \right)\triangleq \left\{ \underbrace{{{g}_{j}}\left( \mathbf{x},\mathbf{y},\mathbf{z} \right)\le 0}_{\text{Inequality constraint}}\text{ }\Leftarrow \left( \mathcal{M}_{n}^{{{\mathcal{T}}_{r}}}\subseteq \tilde{\mathcal{M}}_{n}^{{{\mathcal{T}}_{r}}},\mathcal{M}_{n}^{{{\mathcal{T}}_{r+1}}}\subseteq \tilde{\mathcal{M}}_{n}^{{{\mathcal{T}}_{r+1}}} \right),j=1,2,\ldots ,G \right\}
\end{equation}
where $\mathbf{h}\left( \mathbf{x},\mathbf{y},\mathbf{z} \right)$ is an equation functions family consisting of a set of $H$ equality constraints resulting from \ref{e2.13}, which are named associated resource constraints. $\mathbf{g}\left( \mathbf{x},\mathbf{y},\mathbf{z} \right)$ is a family of $G$ inequality functions that result from the consumption $\mathcal{M}_{n}^{{{\mathcal{T}}_{r}}}$ and $\mathcal{M}_{n}^{{{\mathcal{T}}_{r+1}}}$ of resources in CR ${{\mathcal{T}}_{r}}$ and ${{\mathcal{T}}_{r+1}}$ less than those in SRP $\tilde{\mathcal{M}}_{n}^{{{\mathcal{T}}_{r}}}$ and $\tilde{\mathcal{M}}_{n}^{{{\mathcal{T}}_{r+1}}}$, respectively, which are named limited resource constraints. Variable sets $\mathbf{x}$, $\mathbf{y}$, and $\mathbf{z}$ time, frequency, and computing power resources consumed in IR ${{\mathcal{R}}_{r}}$, respectively.

%
%
%
%
%
%

Next, Thm. \ref{theorem2} formulates the nature of constrained universal resource scheduling.

\begin{theorem}
with the associated resource constraint and the limited resource constraint, the resource cost minimization problem of the client ${{u}_{n}}$ in IR ${{\mathcal{R}}_{r}}\in \mathcal{R}$ remains a convex optimization problem. For any achievable workload ${{N}_{r,n}}\in \left\{ 0,1,2,\ldots ,N_{r,n}^{\max } \right\}$, there is only one optimal resource scheduling strategy $\varpi _{r,n}^{*}\left( {{N}_{r,n}} \right)$ that minimizes the local resource cost function ${{f}_{r,n}}\left( \mathbf{x},\mathbf{y},\mathbf{z} \right)$, where $N_{r,n}^{\max }$ is the maximum transaction volume (MTV) of the client ${{u}_{n}}$ in IR ${{\mathcal{R}}_{r}}$.
\label{theorem2}
\end{theorem}

Proof of Thm. \ref{theorem2} and an explanation about MTV are shown in Appendix \ref{B} and Appendix \ref{C}, respectively. In particular, we define maximum unconstrained transaction volume (MUTV) $N_{r,n}^{\text{unc}}$ as the maximum unconstrained transaction volume with resource constraints, as described in Appendix \ref{C}. From Thm. \ref{theorem1}, when ${{N}_{r,n}}\in \left\{ 0,1,2,\ldots ,N_{r,n}^{\text{unc}} \right\}$, problem \emph{P3} degenerates to problem \emph{P2}, i.e. the unconstrained resource cost minimization problem, when the optimal resource scheduling strategy $\varpi _{r,n}^{*}\left( {{N}_{r,n}} \right)$ can be obtained directly by convex function differentiation. The resource constraints will work when ${{N}_{r,n}}\in \left\{ N_{r,n}^{\text{unc}}+1,N_{r,n}^{\text{unc}}+2,\ldots ,N_{r,n}^{\max } \right\}$.

From Thm. \ref{theorem2}, the optimization problem \emph{P4} is a convex functions with equality and inequality constraints, which works when ${{N}_{r,n}}\in \left\{ N_{r,n}^{\text{unc}}+1,N_{r,n}^{\text{unc}}+2,\ldots ,N_{r,n}^{\max } \right\}$. In this paper, we construct a Lagrange function by the Lagrange-multiplier method, which converts the minimum problem under equality and inequality constraints into an unconstrained optimization problem that satisfies KKT conditions. The Lagrange function constructed for problem \emph{P4} can be described as:

\begin{equation}\label{e4.9}
\mathcal{L}(\mathbf{x},\mathbf{y},\mathbf{z},\mathbf{\mu },\mathbf{\lambda })=\underbrace{{{f}_{r,n}}\left( \mathbf{x},\mathbf{y},\mathbf{z} \right)}_{\text{Optimization function}}+\underbrace{\sum\limits_{{{h}_{i}}\in \mathbf{h}}{{{\mu }_{i}}{{h}_{i}}\left( \mathbf{x},\mathbf{y},\mathbf{z} \right)}}_{\text{Equality constraint}}+\underbrace{\sum\limits_{{{g}_{i}}\in \mathbf{g}}{{{\lambda }_{i}}{{g}_{i}}\left( \mathbf{x},\mathbf{y},\mathbf{z} \right)}}_{\text{Inequality constraint}}
\end{equation}
where $\mathbf{\mu }=\left\{ {{\mu }_{1}},{{\mu }_{2}},\ldots ,{{\mu }_{H}} \right\}$ and $\mathbf{\lambda }=\left\{ {{\lambda }_{1}},{{\lambda }_{2}},\ldots ,{{\lambda }_{G}} \right\}$ are the Lagrange multiplier of the equality and inequality constraints, respectively.

In brief, the proposed M-OURS algorithm is shown in Algo. \ref{algorithm1}.

The M-OURS algorithm proposed in this article adopts a service-oriented resource scheduling approach, which can find the optimal resource scheduling strategy that minimizes the cost of local resources in any achievable workload ${{N}_{r,n}}\in \left\{ 0,1,2,\ldots ,N_{r,n}^{\max } \right\}$. However, it does not emphasize the performance of throughput, latency, et.al, which is significantly different from traditional resource management algorithms for single objective optimization. Baded on the M-OURS algotithm, we can further realize the service-oriented full-lifecycle integration of sensing, communication, and computing (SISCC) mechanism for resource scheduling of MFP service. Regarding the above discussion, we provide the following remark:


\begin{remark}
According to the definition of the economic benefits of the MFP service market and the process of solving the optimal local resource scheduling strategy, it can be seen that our goal is no longer to achieve optimization for a traditional single process and single objective optimization such as maximum throughput, minimum communication delay, or minimum calculation delay, but to minimize the total resource cost for MFP services. Compared to the traditional pursuit of the ultimate single performance indicator, our resource scheduling strategy leans more towards achieving a service-oriented resource balance and complementarity, because we clearly understand that the goal of our resource scheduling is to better meet the comprehensive service requirements of buyers, and a good service experience is not achieved by the ultimate performance of a single physical performance, it must be the comprehensive results of multiple processes and multiple performance indicators. For example, to complete the established MFP service with an achievable workload, if time resources are relatively scarce and expensive, clients tend to choose fewer time resources and more frequency and computational resources, that is, allocate more frequency and computational resources with high latency constraints to compensate for the negative impact of insufficient time resources. On the contrary, when frequency resources or computing power resources are scarce and the unit price is high, clients tend to choose more time resources, that is, use more time resources to compensate for the shortcomings of other resources. Therefore, resource scheduling with the lowest cost of local resources can provide MFP service providers with a highly robust resource guarantee strategy that can automatically adapt to dynamic resource constraints and application requirements.
\label{remark2}
\end{remark}

\section{Experimental results and discussions}

\subsection{Simulation Settings}

\begin{figure}[t]
    \centering
    \subfigure[] {\includegraphics[width=2.5in,angle=0]{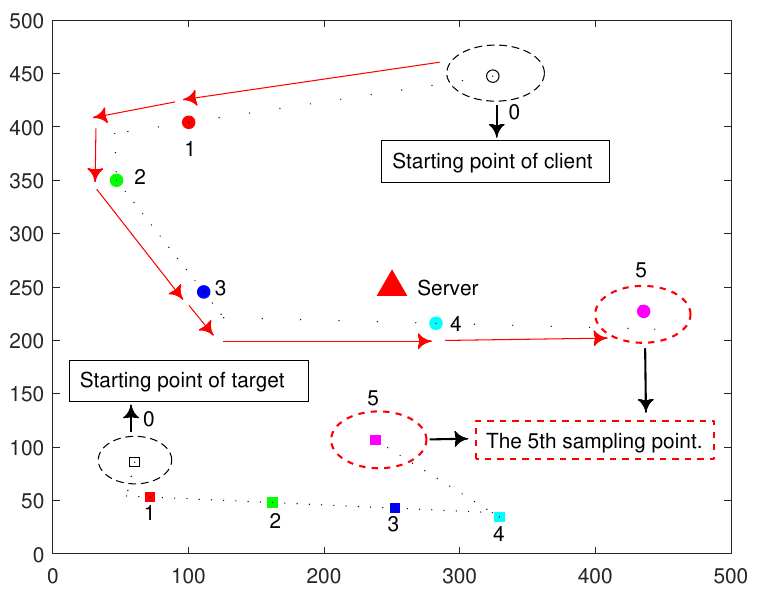}}
    \subfigure[] {\includegraphics[width=2.5in,angle=0]{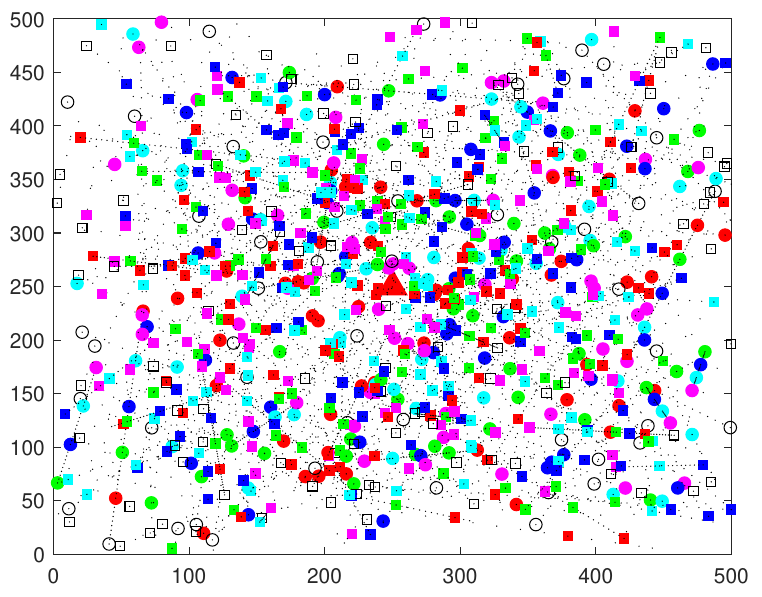}}
    \caption{Simulation scenario and spatial-temporal sampling of intelligent vehicle network.}
    \label{fig5}
\end{figure}

To verify the validity of the proposed service-oriented full-lifecycle integration of sensing, communication, and computing (SISCC) mechanism and the robustness of resource scheduling under the guidance of social welfare, choosing the typical use case of MFP, intelligent vehicle networks, we conduct a lot of simulation and analysis. As shown in Fig. \ref{fig5} (a), consider a square area of $\text{500 }m\text{ }\times \text{ }500\text{ }m$, in which intelligent vehicles act as clients and those unconnected vehicles act as targets. We take CR as a period to sample clients and targets. Fig. \ref{fig5} (b) shows the sampling results of 50 vehicles and 100 targets in 5 consecutive CRs represented by specific colors. Some simulation parameter settings can be seen in Tbl. \ref{tab1}.

\begin{table}
  \caption{SIMULATION PARAMETERS}
  \label{tab1}
  \begin{tabular}{cc}
    \toprule
    PARAMETER&VALUE\\
    \midrule
    Simulation area & $\text{500 }m\text{ }\times \text{ }500\text{ }m$ \\
    Maximum moving speed & 30 $m/s$ \\
    Number of samples & 10000 \\
    Number of servers & 1 \\
    Number of clients & 1:50 \\
    Number of targets & 1:100 \\ 
    Types of targets & 10 \\
    Model size (Global) & 100 $Mbit$ \\
    Model size (Local) & 100 $Mbit$ \\
    Maximum time & 10 $s$ \\
    Maximum computing power & $1 \times {10^6}$ $CPU cycle$ \\
    CPU cycles per sample & 500 $CPU cycle$ \\
    Target-sample ratio (VS) & 10 \\
    Wireless frequency band & 28 $GHz$ \\
    Maximum bandwidth & 400 $MHz$ \\
    Subcarrier spacing & 30 $KHz$ \\
    Transmitting power of RSU & 55 $dBm$ \\
    Transmitting power of client & 26 $dBm$ \\
    Camera frame rate & 20 $fps$ \\
    Size of visual frame & 10 $Mbit$\\
    Visual sensing radius & 50 $m$ \\
    Receiver sensitivity (WS) & -180 $dB$ \\
    Receiver sensitivity (WC) & -115 $dB$ \\
    
  \bottomrule
\end{tabular}
\end{table}

\subsection{Simulation Analysis}
\subsubsection{Benefits of integrated sensing, communication, and computing (ISCC) for MFP service}

\begin{figure}[t]
    \centering
    \subfigure[] {\includegraphics[width=2.5in,angle=0]{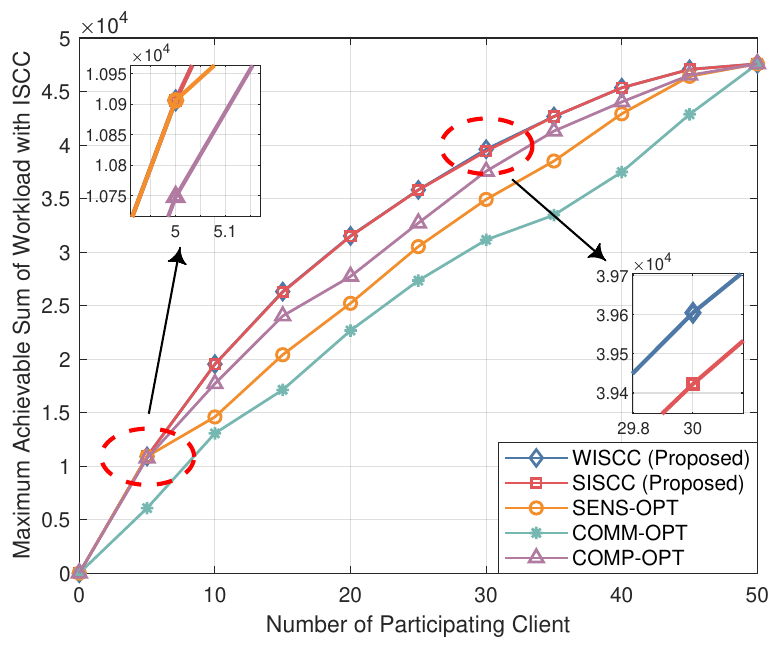}}
    \subfigure[] {\includegraphics[width=2.5in,angle=0]{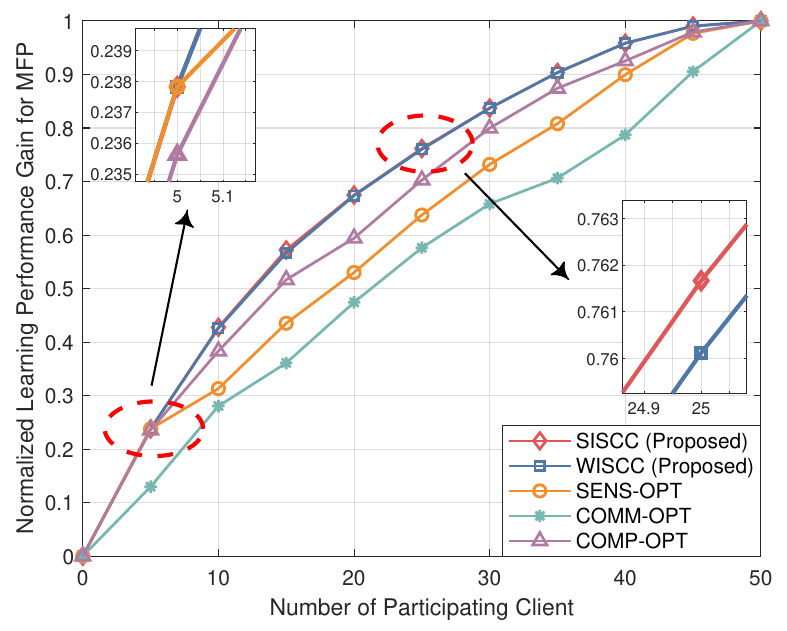}}
    \subfigure[] {\includegraphics[width=2.5in,angle=0]{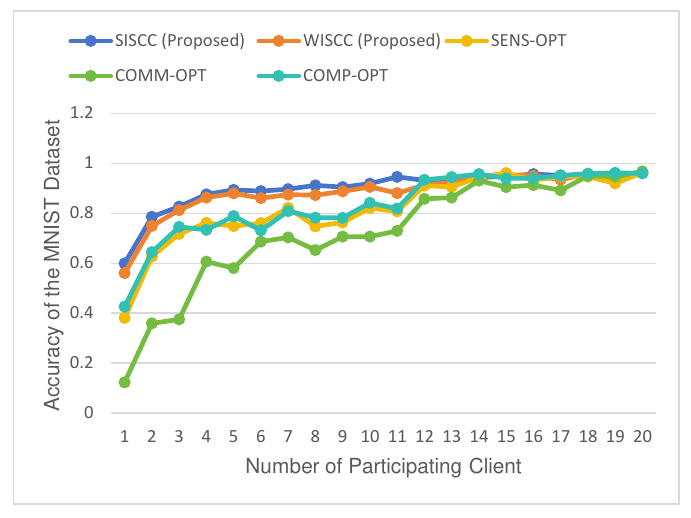}}
    \subfigure[] {\includegraphics[width=2.5in,angle=0]{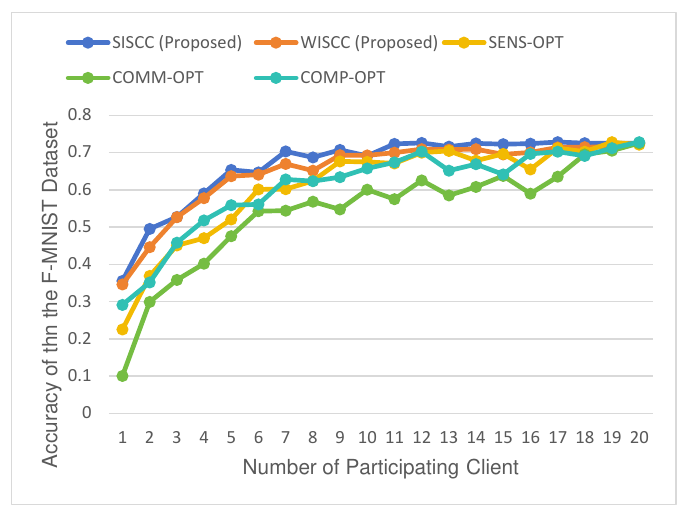}}
    \caption{The validity for MFP of the integrated sensing, communication, and computing (ISCC) mechanisms.}
    \label{fig6}
\end{figure}

The proposed SISCC mechanism is used to guide MFP workload allocation (i.e., client selection) and corresponding resource scheduling, which further considers the impact of quality of data (QoD) based on workload-oriented full-lifecycle integration of sensing, communication, and computing (WISCC). Both SISCC and WISCC are resource scheduling mechanisms with ISCC. We compare them with the sensing-optimal (SENS-OPT), the communication-optimal (COMM-OPT), and the computing-optimal (COMP-OPT) mechanisms, and the results are shown in Fig. \ref{fig6}. As the number of selected working clients (i.e., the clients whose workload is not zero) changes, the obtained maximum achievable sum of workload and the normalized learning performance gain for MFP are shown in Fig. \ref{fig6} (a) and Fig. \ref{fig6} (b), respectively. Furthermore, in Fig. \ref{fig6} (c) and Fig. \ref{fig6} (d), the above simulation results are verified on the real dataset, MNIST and F-MNIST. Fig. \ref{fig6} demonstrates the effectiveness of the proposed WISCC and SISCC mechanisms for MFP service workload and learning performance gain, which can obtain better results than traditional single-process optimization.

\subsubsection{Impact of multimodal data and its sample distribution on MFP service}

\begin{figure}[t]
    \centering
    \subfigure[] {\includegraphics[width=2.5in,angle=0]{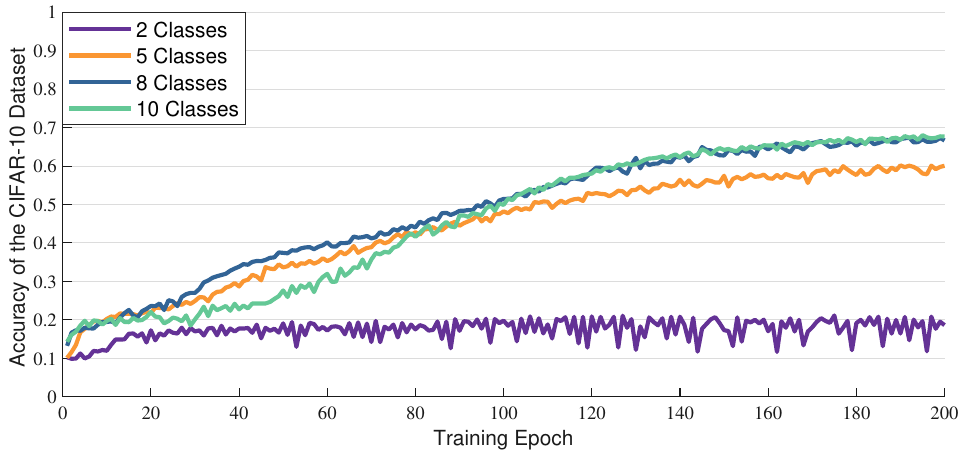}}
    \subfigure[] {\includegraphics[width=2.5in,angle=0]{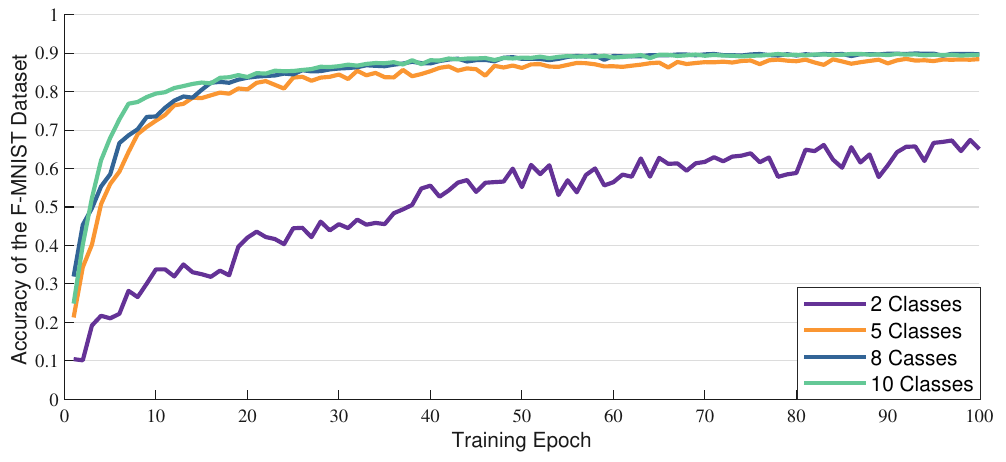}}
    \subfigure[] {\includegraphics[width=2.5in,angle=0]{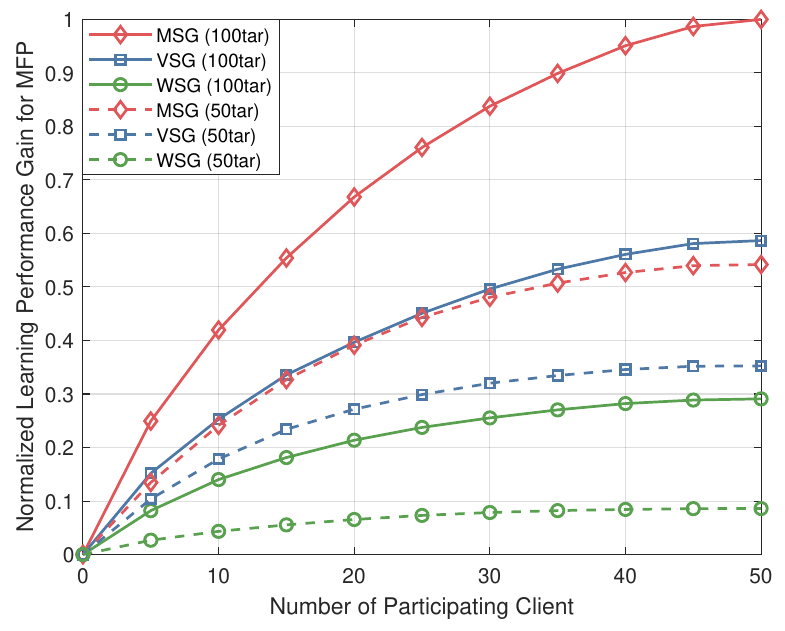}}
    \subfigure[] {\includegraphics[width=2.5in,angle=0]{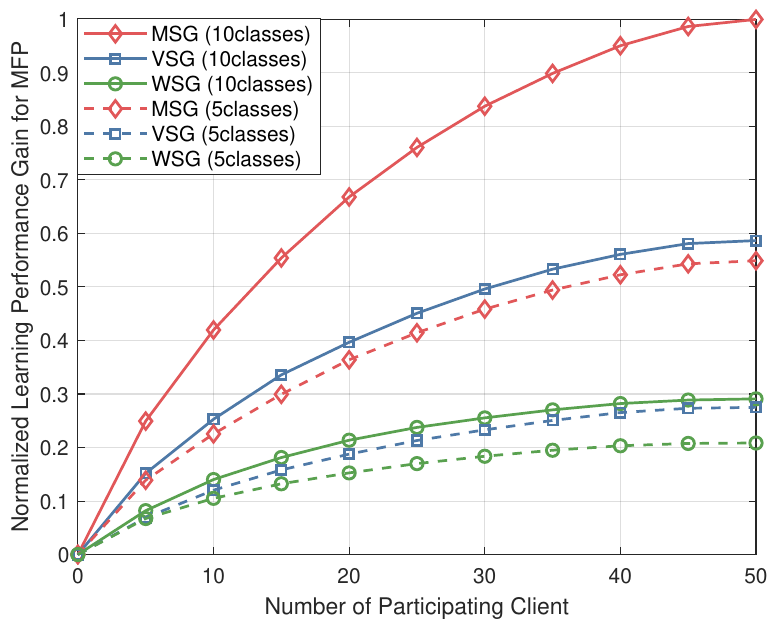}}
    \caption{Impact of multimodal data and target settings.}
    \label{fig7}
\end{figure}

\begin{figure}[htbp]
    \centering
    \subfigure[] {\includegraphics[width=2.0in,angle=0]{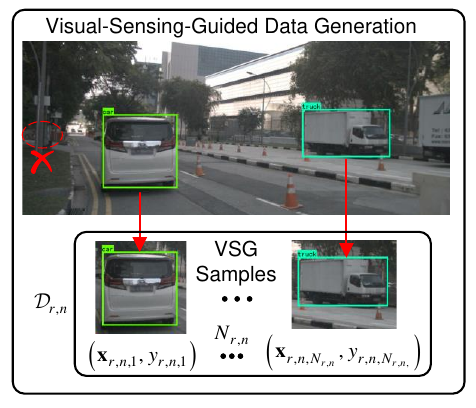}}
    \subfigure[] {\includegraphics[width=2.0in,angle=0]{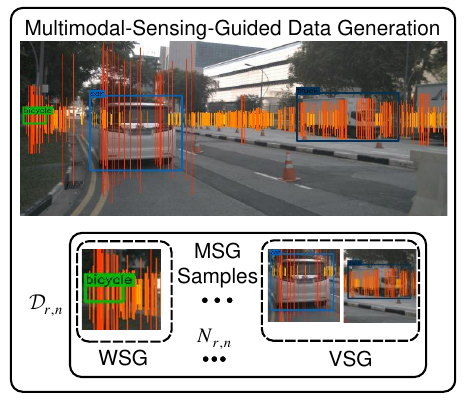}}
    \caption{An example of the progress of VSG- and MSG-based data sample generation.}
    \label{fig8}
\end{figure}

First, Fig. \ref{fig7} shows that compared to single-modal data generation, i.e., visual-sensing-guided (VSG) and wireless-sensing-guided (WSG) acquisition of training samples, the multimodal-sensing-guided (MSG) data generation can achieve better learning performance gain for MFP. In particular, MSG is not a simple superposition of VSG and WSG, which can improve the 	QoD and realize ``making one plus one bigger than two'' by reducing the difference between local samples and global data through sample distribution complementarity. In addition, the number and variety of targets sensed in the environment can also have an impact on the quality of service (QoS), as it directly affects the generation of multimodal data samples. In addition, we also train in the F-MNIST and CIFAR-10 datasets to show the impact of the number of data classes, which is consistent with the simulation results in this paper \cite{mcmahan2017communication,xiao2017fashion}. Furthermore, we demonstrate the effectiveness of multimodal data generation on the nuScenes dataset \cite{caesar2020nuscenes}, as shown in Fig. \ref{fig8} and  Tbl. \ref{tab2}. The CRF-Net network \cite{nobis2019deep} based on multimodal sensing improves the performance on the mean Average Precision (mAP) compared with the pure-vision sensing based on YOLOv4 and RetinaNet \cite{bochkovskiy2020yolov4, lin2017focal}.

\begin{table*}[t]
\footnotesize
\caption{Accuracy comparison of diverse sensing strategies on the nuScenes dataset.}
\label{tab2}
\renewcommand\arraystretch{1.5}
\begin{tabular}{|c|c|c|c|c|c|}
\hline
\textbf{Network}  &\textbf{Dataset}  &\textbf{mAP}  \\ \hline
\textbf{YOLOv4} & nuScenes & 35.32\% \\ \hline
\textbf{RetinaNet} & nuScenes & 43.47\% \\ \hline
\textbf{CRF-Net} & nuScenes & 43.95\% \\ \hline
\end{tabular}
\end{table*}

\subsubsection{Effectiveness of SISCC mechanism on client selection }

\begin{figure}[t]
 \centerline{\includegraphics[width=3.5in]{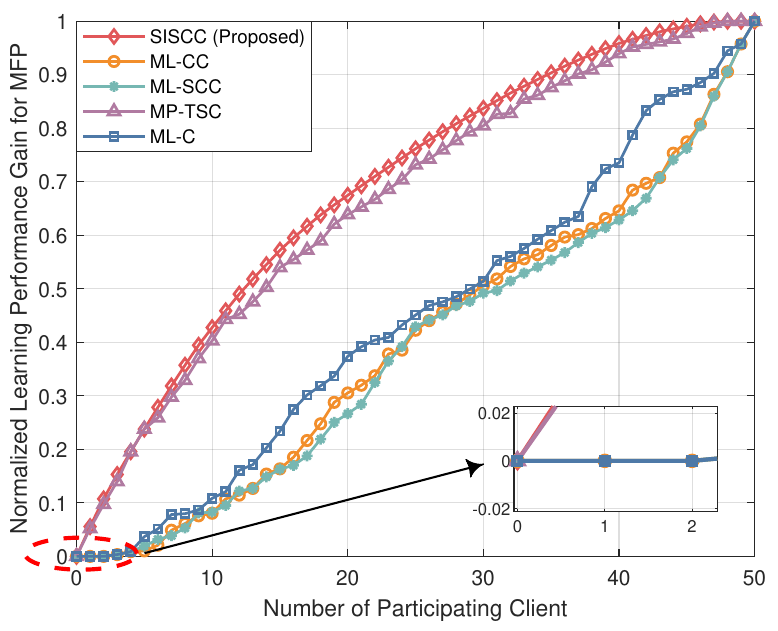}}
 \caption{Impact of client selection policy on MFP service.}
 \centering
 \label{fig9}
\end{figure}

Fig. \ref{fig9} shows the performance of the proposed SISCC mechanism on client selection, which proves that the proposed strategy can achieve the best MFP learning performance gain with the same number of working clients. The benchmarks include: 1) client selection with the minimum latency in communication (ML-C); 2) client selection with the minimum sum of latency in communication and computing (ML-CC). Optimization to minimize communication and training delay is the major route in current wireless FL optimization. The modeling of ML-CC part-referenced the idea of communication and computing delay minimization in \cite{chen2020convergence}; 3) client selection with the minimum sum of latency in sensing, communication, and computing (ML-SCC). Based on minimizing the communication and computation delay, and further considering the sensing delay, ML-SCC can realize the minimization of the sum latency resulting from sensing, communication, and computing. The modeling of the ML-SCC algorithm part-referenced the idea of time modeling in \cite{tang2023integrated}; 4) client selection with the maximum product of the number of sensed targets and the sensing capacity (MP-TSC). Authors in \cite{saputra2021dynamic} successively consider the impact of location and the data information in the client selection problem. The modeling of the MP-TSC algorithm part-referenced the problem modeling in \cite{saputra2021dynamic}.

\subsubsection{ Impact of multi-domain resources on MFP service}

\begin{figure}[t]
    \centering
    \subfigure[] {\includegraphics[width=1.8in,angle=0]{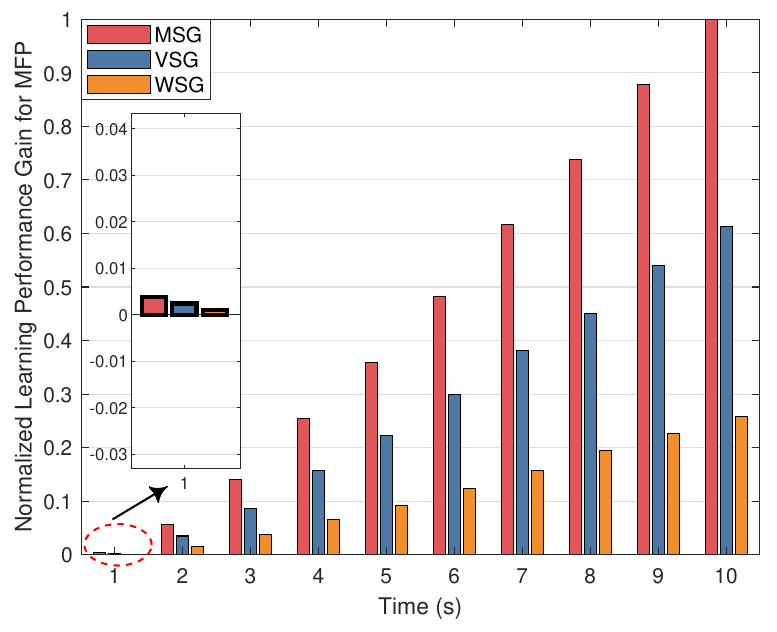}}
    \subfigure[] {\includegraphics[width=1.8in,angle=0]{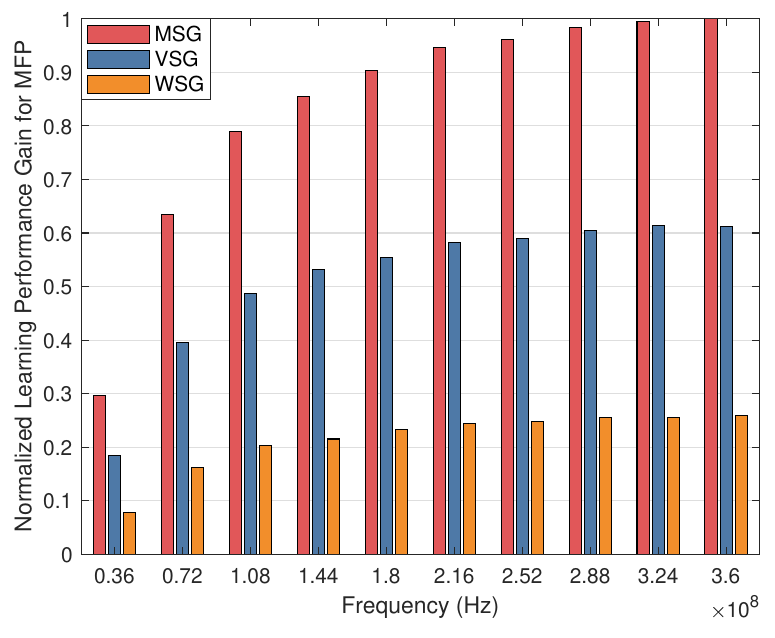}}
    \subfigure[] {\includegraphics[width=1.8in,angle=0]{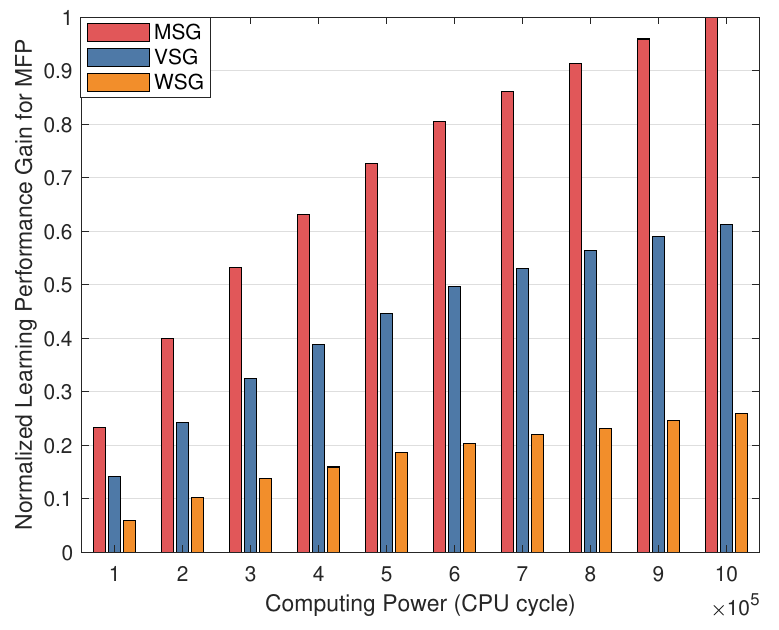}}
    \caption{Representation of the relationship between multi-domain resources and MFP service (2D).}
    \label{fig10}
\end{figure}

\begin{figure}[t]
    \centering
    \subfigure[] {\includegraphics[width=1.5in,angle=0]{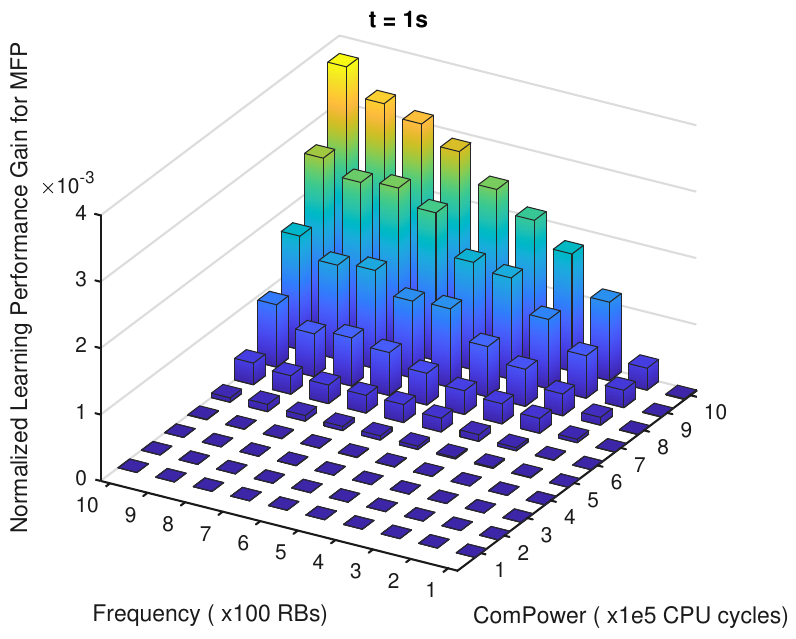}}
    \subfigure[] {\includegraphics[width=1.5in,angle=0]{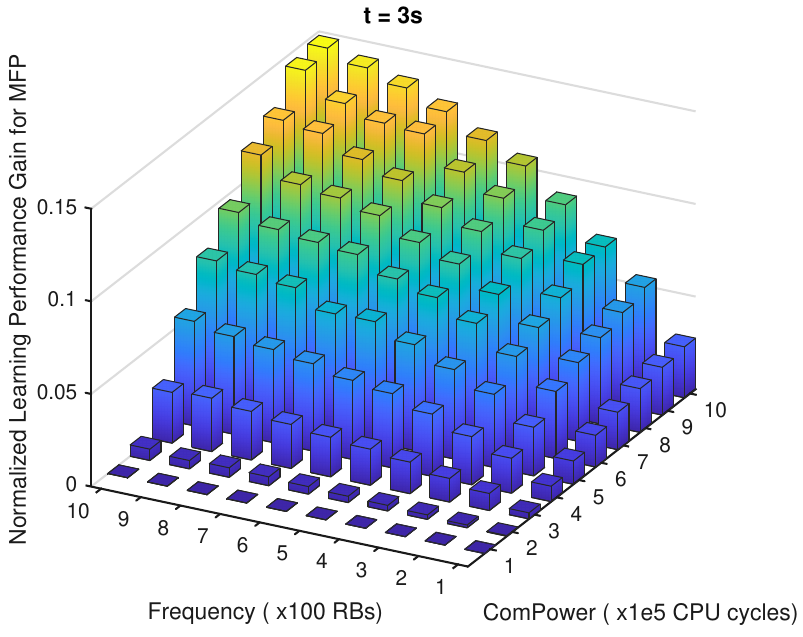}}
    \subfigure[] {\includegraphics[width=1.5in,angle=0]{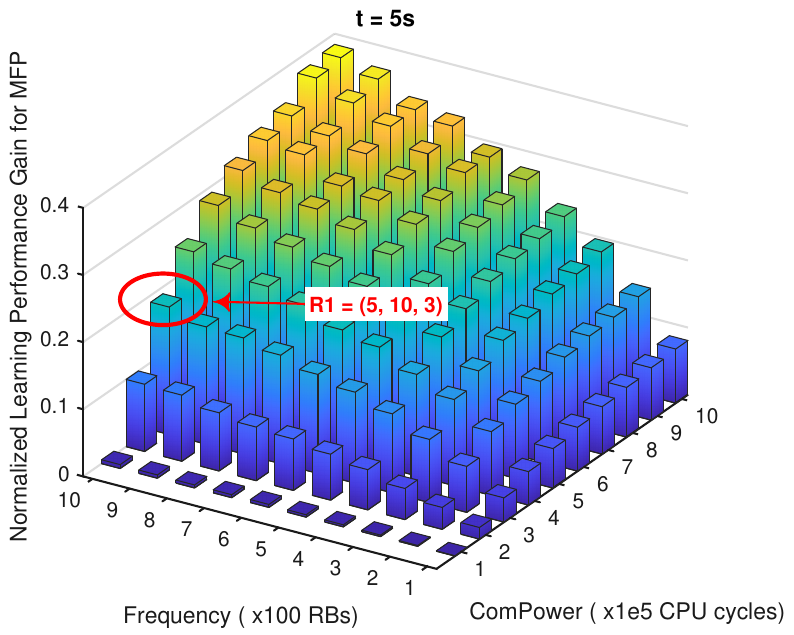}}
    \subfigure[] {\includegraphics[width=1.5in,angle=0]{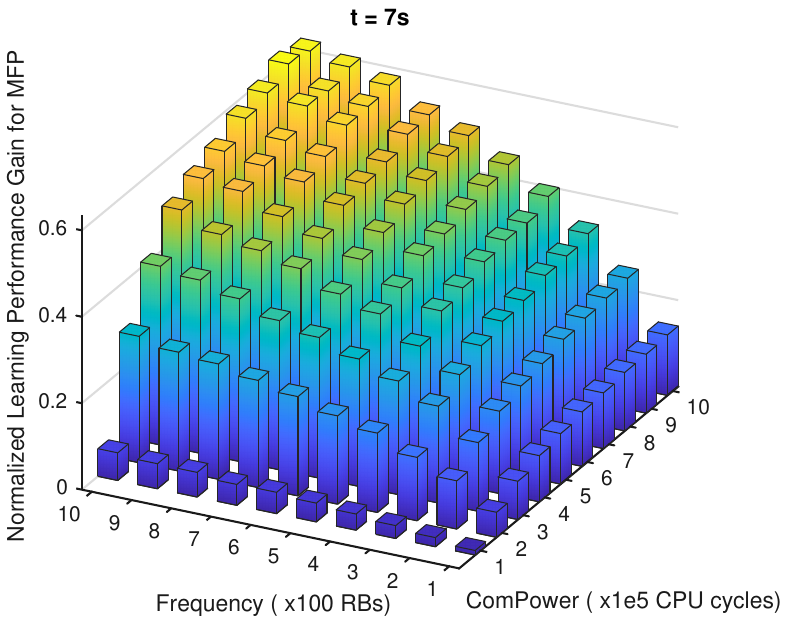}}
    \subfigure[] {\includegraphics[width=1.5in,angle=0]{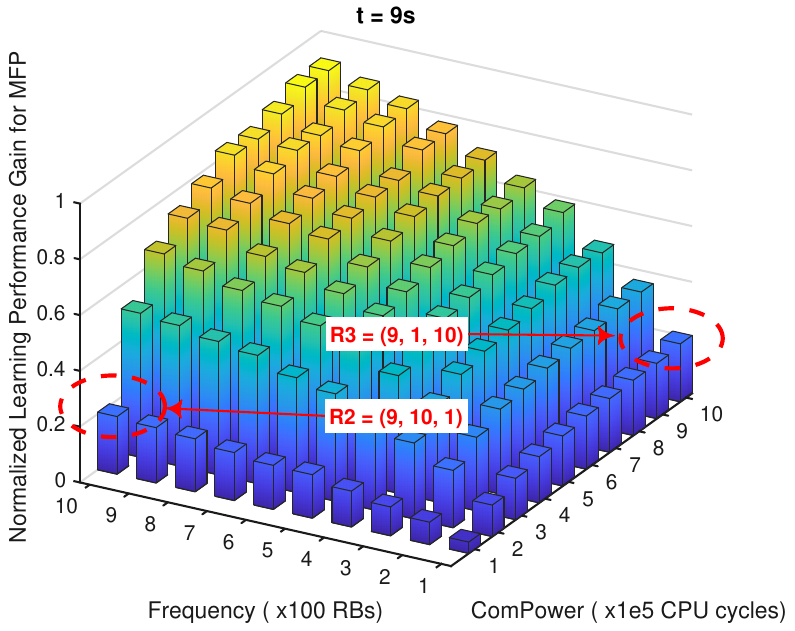}}
    \subfigure[] {\includegraphics[width=1.5in,angle=0]{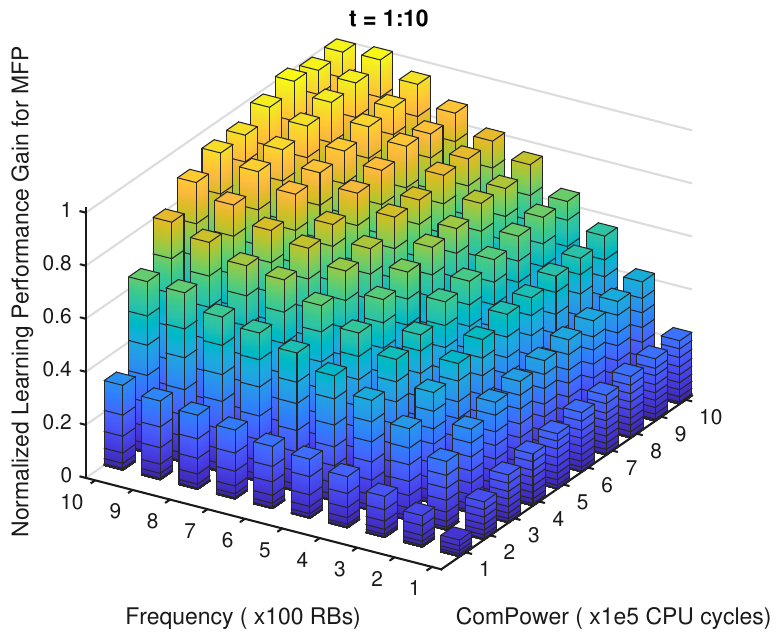}}
    \caption{Representation of the relationship between multi-domain resources and MFP service (3D).}
    \label{fig11}
\end{figure}

Fig. \ref{fig10} shows the impact of the multi-domain resources shared by sensing, communication, and computing, including time, frequency, and computing power, on the learning performance of MFP services. Furthermore, Fig. \ref{fig11} explores the 3D image of the relationship between MFP service performance and frequency and computing power resources. Ulteriorly, if the time variable is further taken into account, we can obtain the changing trend of MFP normalized learning performance gain with the three multi-domain resources composed of time, frequency, and computing power, which is shown in Fig. \ref{fig11} (f). From the above results, we can find that the performance of MFP service is more sensitive to time resources, followed by computing power, and relatively insensitive to frequency resources, under the current situation of $1s$, $100 RB$, and $1\times {{10}^{5}} CPU cycles$ as the quantitative units of time, frequency, and computing power resources.

\subsubsection{Effectiveness and robustness of MFP service transaction mechanism.}

\begin{figure}[t]
    \centering
    \subfigure[] {\includegraphics[width=1.5in,angle=0]{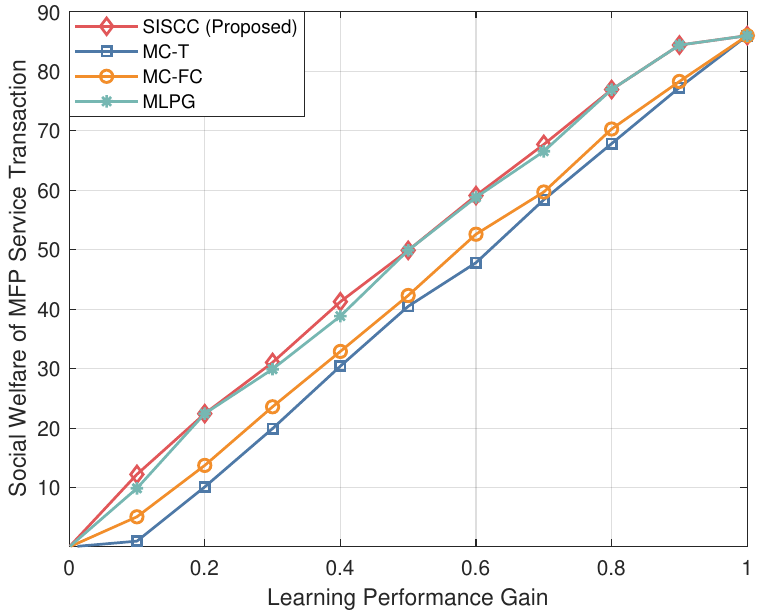}}
    \subfigure[] {\includegraphics[width=1.5in,angle=0]{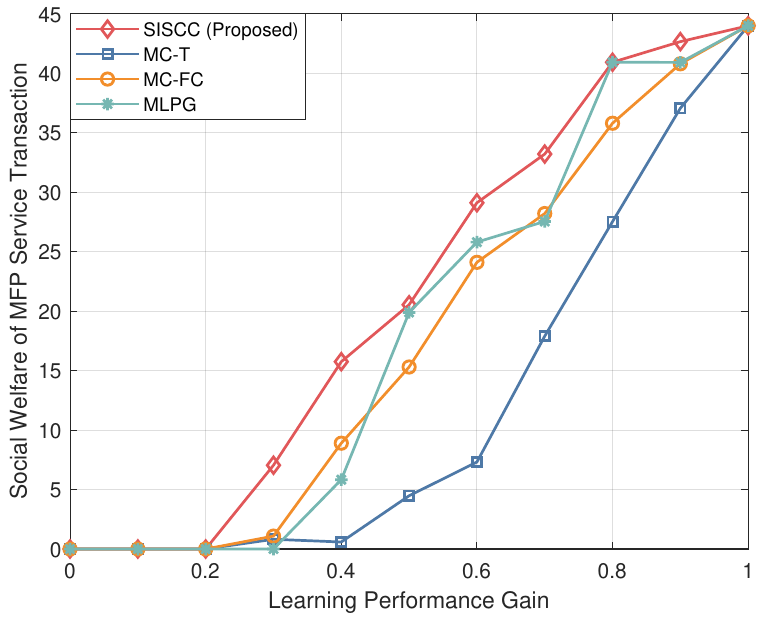}}
    \subfigure[] {\includegraphics[width=1.5in,angle=0]{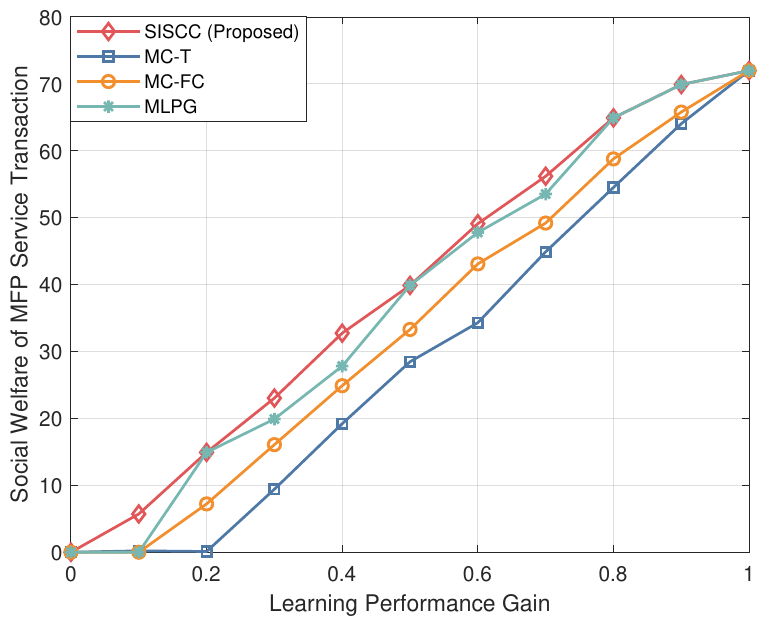}}
    \subfigure[] {\includegraphics[width=1.5in,angle=0]{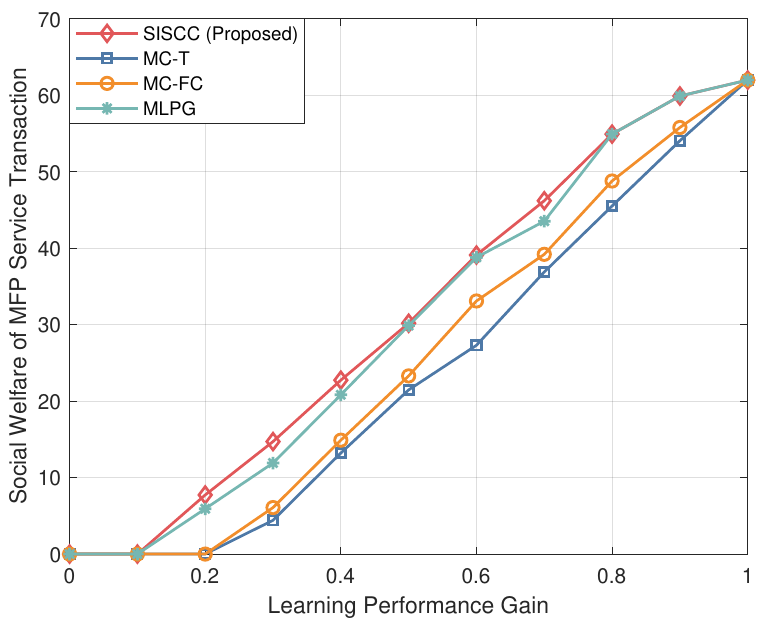}}
    \subfigure[] {\includegraphics[width=1.5in,angle=0]{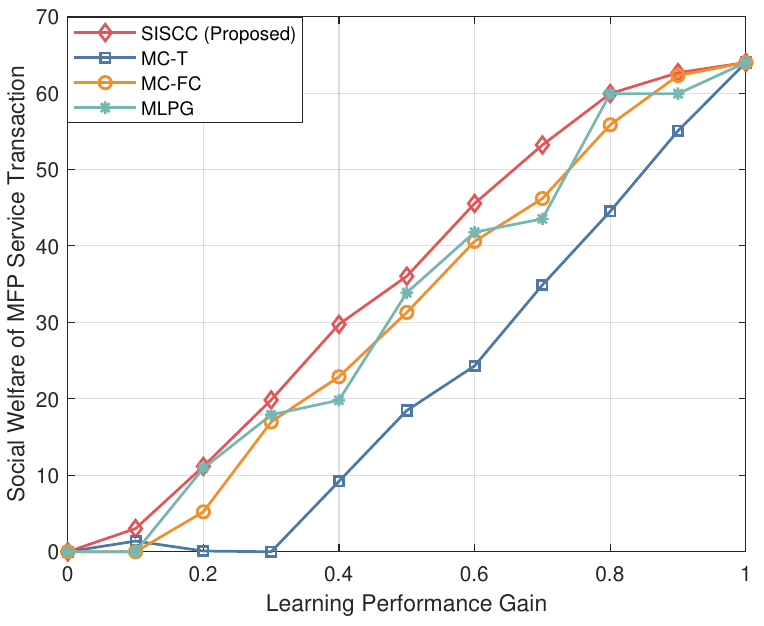}}
    \subfigure[] {\includegraphics[width=1.5in,angle=0]{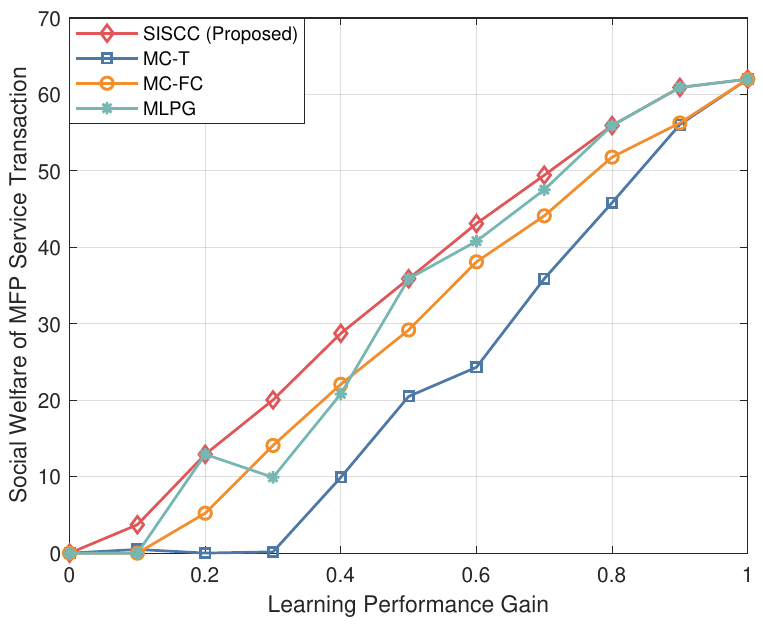}}
    \caption{Impact of local resource scheduling policy on MFP service.}
    \label{fig12}
\end{figure}

Fig. \ref{fig12} shows the effectiveness and robustness of the SISCC-guided local resource scheduling strategy with maximum social welfare. Fig. \ref{fig12} (a-b) shows that in the resource-sufficient scenario (where the normalized amount of time, frequency, and computing power resource is defined as $1:1:1$) and the resource-constricted scenario (the corresponding time, frequency, and computing power resource is $1/4:1/4:1/4$), the changing trend of the social welfare of the MFP service market with the learning performance gain demand of the upper application. First, according to the previous discussion and the results in Fig. \ref{fig11}, it can be known that the client ${{u}_{n}}\in \mathcal{U}$ can obtain a certain MTV $N_{r,n}^{\max }$ and the corresponding MFP learning performance gain with the available resources. Therefore, in the resource-constricted scenario, the model gain that the client can obtain will be reduced. However, compared with benchmarks, the proposed mechanism can still obtain the maximum social welfare due to the implementation of unified resource scheduling with ISCC. Meanwhile, taking the resource combination $1/2:1/2:1/2$ in Fig. \ref{fig12} (c) as the baseline resource stock, we show the social welfare under the time-resource-shortage scenario ($1/4:1/2:1/2$), frequency-resource-shortage scenario ($1/2:1/4:1/2$) and computing power resource shortage scenario ($1/2:1/2:1/4$) in Fig. \ref{fig12} (d-f). From the above results, we can see that the proposed SISCC-guided resource scheduling strategy can obtain the maximum social welfare. Since the gross profit obtained from the upper application and the learning performance gain are positively correlated, the above results imply that we achieved the lowest resource cost. In addition, Fig. \ref{fig12} shows that the changing trend of the curve of our proposed strategy is more stable and does not produce large fluctuations with the dynamic change of resource constraints. 

The benchmarks for comparison include 1) resource scheduling with the minimum cost of time (MC-T). The delay cost is an important object for FL optimization. The modeling of the MC-T algorithm part-referenced the time consumption model in \cite{cheng2023cheese}; 2) resource scheduling with the minimum cost of frequency and computing power (MC-FC), which part-referenced the idea of resource cost modeling in \cite{liu2022resource}; 3) resource scheduling with the maximum learning performance gain (MLPG), which only focuses on the gross profit generated by learning performance gain, while the optimization of resource cost is ignored.

\begin{figure}[t]
 \centerline{\includegraphics[width=5.5in]{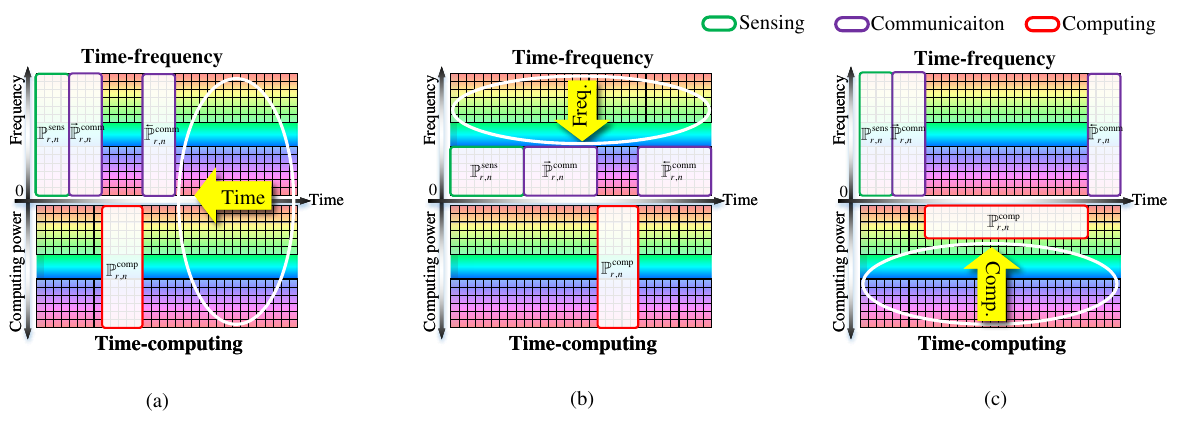}}
 \caption{The regularity of minimum resource costs under dynamic resource constraints.}
 \centering
 \label{fig13}
\end{figure}

In particular, to facilitate a clear and intuitive understanding of the robustness of the proposed SISCC-based MFP service transaction mechanism, we give an intuitive presentation of its tendency toward resource cost minimization, as shown in Fig. \ref{fig13}. Fig. \ref{fig13} (a) indicates that when time resources are limited or with stringent specifications on latency, the proposed algorithm can automatically optimize the resource scheduling strategy in the direction of minimizing time resource consumption. The other two cases are similar. In Fig. \ref{fig11}, we give another simple and intuitive example. When the learning performance gain requirement of the MFP service is ${{\theta }_{0}}=0.2$, we have a variety of resource scheduling strategies for the combination of time, frequency, and computing power, such as $\left( 5,10,3 \right)$, $\left( 9,1,10 \right)$, and$\left( 9,10,1 \right)$. Under the circumstances, we can flexibly select the resource combination that minimizes the total resource cost according to the resource price and available resource stock in the current scenario, that is, to realize the maximum social welfare of MFP service transactions.

\section{Conclusion}

This paper studies the problem of optimal resource scheduling for MFP services. First, the process of sensing-based multimodel data sample generation is incorporated into the optimization of FL, and a resource consumption minimization problem for learning performance gain of MFP models is established. Then, utilizing the workload of multimodal data samples as a link between data generation and consumption, the learning gain of the MFP model and the indirect mathematical model with multidomain physical resources are established to achieve the bottom-level physical resource scheduling for upper-level services. Further, we redefine the optimization function as a cost-efficient social welfare maximization problem in the MFP service market and achieve the local minimum resource cost by the resource scheduling and cost analysis of process-decoupled SRPs. Comprehensive simulation results illustrate the proposed mechanism is efficient and robust.


%


\bibliographystyle{ieeetr}
\bibliography{ISCC-MFP}

\appendix

\clearpage
\section{Proof of Theorem 1.}
\label{A}

Due to the independence of SRP, it can be obtained that:

\begin{equation}\label{b1.1}
\underset{\varpi _{r,n}^{*}}{\mathop{\text{arg min:}}}\,{{c}_{r,n}}\left( {{N}_{r,n}} \right)\Leftrightarrow \underset{\varpi _{r,n,r}^{*}}{\mathop{\text{arg min:}}}\,c_{r,n}^{\text{d}\downarrow }\left( {{N}_{r,n}} \right)+\underset{\varpi _{r,n,r}^{*}}{\mathop{\text{arg min:}}}\,c_{r,n}^{\text{d}\uparrow }\left( {{N}_{r,n}} \right)
\end{equation}

Therefore, the original problem is equivalent to proving that the resource cost function $c_{r,n}^{\text{d}\downarrow }\left( {{N}_{r,n}} \right)$ of the data generation $\mathbb{P}_{r,n}^{\text{d}\downarrow }$ in ${{\mathcal{T}}_{r}}$ and the resource cost function $c_{r,n}^{\text{d}\uparrow }\left( {{N}_{r,n}} \right)$ of the data consumption $\mathbb{P}_{r,n}^{\text{d}\uparrow }$ in ${{\mathcal{T}}_{r+1}}$ are both strongly convex functions, with one and only one minimum point.

Below we will provide proof separately.

\begin{enumerate}[1)]
\item For data generation $\mathbb{P}_{r,n}^{\text{d}\downarrow }$ in ${{\mathcal{T}}_{r}}$:
\end{enumerate}

The relationship between the workload ${{N}_{r,n}}$ of $\mathbb{P}_{r,n}^{\text{d}\downarrow }$ and resources can be expressed as:

\begin{equation}\label{b1.2}
{{N}_{r,n}}\approx {{\lambda }_{r,n}}N_{r,n}^{\text{fre}} =\left\lfloor \underbrace{\rho _{r,n}^{\text{tar}}{{S}_{\text{vs}}}\delta _{r,n}^{\text{vs}}T_{r,n}^{\text{vs}}{{f}_{\text{vs}}}}_{\text{visual sensing}}+\underbrace{\rho _{r,n}^{\text{tar}}\frac{T_{r,n}^{\text{ws}}}{T_{r,n}^{\text{vs}}}\left( {{S}_{\text{ws}}}-{{S}_{\text{vs}}} \right){{\varepsilon }_{\text{ws}}}B_{r,n}^{\text{ws}}{{\log }_{2}}\left( 1+\frac{P_{r,n}^{\text{ws}}{{{\tilde{h}}}_{r,n}}}{{{N}_{0}}} \right)T_{r,n}^{\text{vs}}{{f}_{\text{vs}}}}_{\text{wireless sensing}} \right\rfloor \\
\end{equation}


Define the status attribute:

%
%

\begin{equation}\label{b1.3}
{{a}_{r,n}}=\rho _{r,n}^{\text{tar}}{{S}_{\text{vs}}}\delta _{r,n}^{\text{vs}}{{f}_{\text{vs}}},{{b}_{r,n}}=\rho _{r,n}^{\text{tar}}\left( {{S}_{\text{ws}}}-{{S}_{\text{vs}}} \right){{\varepsilon }_{\text{ws}}}{{\log }_{2}}\left( 1+\frac{P_{r,n}^{\text{ws}}{{{\tilde{h}}}_{r,n}}}{{{N}_{0}}} \right){{f}_{\text{vs}}}
\end{equation}
which varies with the clients' state in different IRs, but can be considered constant in each IR, then:

\begin{equation}\label{b1.5}
{{N}_{r,n}}=\left\lfloor \underbrace{{{a}_{r,n}}T_{r,n}^{\text{vs}}}_{\text{visual sensing}}+\underbrace{{{b}_{r,n}}T_{r,n}^{\text{ws}}B_{r,n}^{\text{ws}}}_{\text{wireless sensing}} \right\rfloor 
\end{equation}

Due to $T_{r,n}^{\text{ws}}\le T_{r,n}^{\text{vs}}$ and the need for wireless sensing to synchronize with visual sensing, based on the definition of resource cost, only fees for additional $T_{r,n}^{\text{vs}}$ need to be paid.

Denote $x=T_{r,n}^{\text{vs}}$, $y=B_{r,n}^{\text{ws}}$, $z=T_{r,n}^{\text{ws}}$, then:

\begin{equation}\label{b1.6}
{{N}_{r,n}}={{a}_{r,n}}x+{{b}_{r,n}}zy
\end{equation}

The cost function $c_{r,n}^{\text{d}\downarrow }$ for data generation $\mathbb{P}_{r,n}^{\text{d}\downarrow }$:

\begin{equation}\label{b1.7}
c_{r,n}^{\text{d}\downarrow }=T_{r,n}^{\text{vs}}{{\Delta }_{\text{t}}}+B_{r,n}^{\text{ws}}{{\Delta }_{\text{b}}}=x{{\Delta }_{\text{t}}}+y{{\Delta }_{\text{b}}}=x{{\Delta }_{\text{t}}}+\left( \frac{{{N}_{r,n}}-{{a}_{r,n}}x}{{{b}_{r,n}}z} \right){{\Delta }_{\text{b}}}
\end{equation}
%
%

From \ref{b1.7}, it can be seen that $z\uparrow \to \left( c_{r,n}^{\text{d}\downarrow } \right)\downarrow$, when the maximum value $z=x$ is taken, we can get:

\begin{equation}\label{b1.9}
c_{r,n}^{\text{d}\downarrow }=x{{\Delta }_{\text{t}}}+\frac{{{N}_{r,n}}{{\Delta }_{\text{b}}}}{{{b}_{r,n}}}\frac{1}{x}-\frac{{{a}_{r,n}}{{\Delta }_{\text{b}}}}{{{b}_{r,n}}}
\end{equation}


Furthermore,  $c_{r,n}^{\text{d}\downarrow }$ takes the derivative of $x$ and sets it to 0 to obtain a unique solution ${{x}_{0}}$. Substitute ${{x}_{0}}$ and ${{z}_{0}}={{x}_{0}}$ into \ref{b1.6} to obtain the local resource scheduling strategy $\varpi _{r,n,r}^{*}\left( {{N}_{r,n}} \right)$ that minimizes the resource cost of data generation $\mathbb{P}_{r,n}^{\text{d}\downarrow }$ with workload ${{N}_{r,n}}$:

\begin{equation}\label{b1.11}
\underset{\varpi _{r,n,r}^{*}}{\mathop{\text{arg min:}}}\,c_{r,n}^{\text{d}\downarrow }\left( {{N}_{r,n}} \right)=\left\{ {{x}_{0}},{{y}_{0}},{{z}_{0}} \right\}=\left\{ \sqrt{\frac{{{N}_{r,n}}{{\Delta }_{\text{b}}}}{{{b}_{r,n}}{{\Delta }_{\text{t}}}}},\frac{1}{{{b}_{r,n}}}\left( \sqrt{\frac{{{N}_{r,n}}{{b}_{r,n}}{{\Delta }_{\text{t}}}}{{{\Delta }_{\text{b}}}}}-{{a}_{r,n}} \right),\sqrt{\frac{{{N}_{r,n}}{{\Delta }_{\text{b}}}}{{{b}_{r,n}}{{\Delta }_{\text{t}}}}} \right\}
\end{equation}


Therefore, there is only one proven minimum value for the resource cost of data generation $\mathbb{P}_{r,n}^{\text{d}\downarrow }$:

\begin{equation}\label{b1.12}
\text{minimize}:c_{r,n}^{\text{d}\downarrow }\left( {{N}_{r,n}} \right)=\sqrt{\frac{{{N}_{r,n}}{{\Delta }_{\text{b}}}}{{{b}_{r,n}}{{\Delta }_{\text{t}}}}}{{\Delta }_{\text{t}}}+\frac{1}{{{b}_{r,n}}}\left( \sqrt{\frac{{{N}_{r,n}}{{b}_{r,n}}{{\Delta }_{\text{t}}}}{{{\Delta }_{\text{b}}}}}-{{a}_{r,n}} \right){{\Delta }_{\text{b}}}
\end{equation}

\begin{enumerate}[2)]
\item For data consumption $\mathbb{P}_{r,n}^{\text{d}\uparrow }$ in ${{\mathcal{T}}_{r+1}}$:
\end{enumerate}

Because data consumption $\mathbb{P}_{r,n}^{\text{d}\uparrow }$ consists of strictly serially executed  $\vec{\mathbb{P}}_{r,n}^{\text{comm}}$, $\mathbb{P}_{r,n}^{\text{comp}}$, and $\overset{\scriptscriptstyle\leftarrow}{\mathbb{P}}_{r,n}^{\text{comm}}$, and ${{D}_{\text{w}}}$, the amount of transmitted data for $\vec{\mathbb{P}}_{r,n}^{\text{comm}}$ and $\overset{\scriptscriptstyle\leftarrow}{\mathbb{P}}_{r,n}^{\text{comm}}$, is constant and independent from workload, the minimum of resource cost $c_{r,n}^{\text{d}\uparrow }$ in the data consumption $\mathbb{P}_{r,n}^{\text{d}\uparrow }$ can be defined as the problem of finding the minimum resource cost of subprocesses, $\vec{\mathbb{P}}_{r,n}^{\text{comm}}$, $\mathbb{P}_{r,n}^{\text{comp}}$, and $\overset{\scriptscriptstyle\leftarrow}{\mathbb{P}}_{r,n}^{\text{comm}}$:

\begin{equation}\label{b1.13}
\underset{\varpi _{r,n,r+1}^{*}}{\mathop{\text{arg min:}}}\,c_{r,n}^{\text{d}\uparrow }\left( {{N}_{r,n}} \right)\Leftrightarrow \underset{\varpi _{r,n,r+1}^{*}|\vec{\mathbb{P}}_{r,n}^{\text{comm}}}{\mathop{\text{arg min:}}}\,\vec{c}_{r,n}^{\text{comm}}\left( {{N}_{r,n}} \right)+\underset{\varpi _{r,n,r+1}^{*}|\mathbb{P}_{r,n}^{\text{comp}}}{\mathop{\text{arg min:}}}\,c_{r,n}^{\text{comp}}\left( {{N}_{r,n}} \right)+\underset{\varpi _{r,n,r+1}^{*}|\overset{\scriptscriptstyle\leftarrow}{\mathbb{P}}_{r,n}^{\text{comm}}}{\mathop{\text{arg min:}}}\,\overset{\scriptscriptstyle\leftarrow}{c}_{r,n}^{\text{comm}}\left( {{N}_{r,n}} \right) \\ 
\end{equation}

Next, by solving the three constituent elements of the above equation, it can be proven that there is only one local resource scheduling strategy that minimizes the resource cost $c_{r,n}^{\text{d}\uparrow }\left( {{N}_{r,n}} \right)$ of data consumption $\mathbb{P}_{r,n}^{\text{d}\uparrow }$ with workload ${{N}_{r,n}}$.

\begin{equation}\label{b1.14}
\varpi _{r,n,r+1}^{*}\left( {{N}_{r,n}} \right)=\sum\limits_{\mathbb{P}\in \mathbb{P}_{r,n}^{\text{d}\uparrow }}{\varpi _{r,n,r+1}^{*}|\mathbb{P}}
\end{equation}

The proof process is similar to the above and will not be repeated here.

From the above analysis, it can be proved that the resource cost minimization problem for clients ${{u}_{n}}$ in IR ${{\mathcal{R}}_{r}}\in \mathcal{R}$ with unconstrained conditions is a strongly convex optimization problem. For any given workload ${{N}_{r,n}}$, there is only one optimal resource scheduling policy $\varpi _{r,n}^{*}\left( {{N}_{r,n}} \right)$ that minimizes the cost of local resources.

\section{The Proof of Theorem 2 and the explanation of MTV.}
\label{B}

Although SRP $\tilde{\mathcal{M}}_{n}^{{{\mathcal{T}}_{r}}}=\left\{ \mathbf{\tilde{b}}_{r,n}^{\mathbf{t}},\mathbf{\tilde{f}}_{r,n}^{\mathbf{t}} \right\}$ and $\tilde{\mathcal{M}}_{n}^{{{\mathcal{T}}_{r+1}}}=\left\{ \mathbf{\tilde{b}}_{r+1,n}^{\mathbf{t}},\mathbf{\tilde{f}}_{r+1,n}^{\mathbf{t}} \right\}$ of ${{\mathcal{T}}_{r}}$ and ${{\mathcal{T}}_{r+1}}$ are independent, the existence of a family of equality-constrained functions $\mathbf{h}\left( \mathbf{x},\mathbf{y},\mathbf{z} \right)$ requires joint optimization of two CRs in Lagrange functions. If the family of equality-constrained functions $\mathbf{h}\left( \mathbf{x},\mathbf{y},\mathbf{z} \right)$ is ignored and independent optimization is performed on the resource scheduling ${{\mathcal{T}}_{r}}$ and ${{\mathcal{T}}_{r+1}}$ with limited resource constraints, the number of valid solutions will be less than the number of solutions with equality constraints. 

Therefore, we only need to prove that independent optimization has one or no optimal solution, that is, joint optimization with equality constraints can also have at most one or no optimal solution.

Next, we solve the problem of minimizing the resource cost of data generation $\mathbb{P}_{r,n}^{\text{d}\downarrow }$ in ${{\mathcal{T}}_{r}}$ and data consumption $\mathbb{P}_{r,n}^{\text{d}\uparrow }$ in ${{\mathcal{T}}_{r+1}}$ under the constraints of limited resources based on Lagrange multipliers. Among them, some definitions extend the contents of Appendix \ref{B}.

\begin{enumerate}[1)]
\item For data generation $\mathbb{P}_{r,n}^{\text{d}\downarrow }$ in ${{\mathcal{T}}_{r}}$:
\end{enumerate}

\begin{equation}\label{c1.4}
\text{minimize: }f_{r,n}^{\text{d}\downarrow }\left( x,y \right)=x{{\Delta }_{\text{t}}}+y{{\Delta }_{\text{b}}}
\end{equation}
S.T.

\begin{equation}\label{c1.5}
{{h}_{1}}\left( x,y \right)={{N}_{r,n}}-{{a}_{r,n}}x-{{b}_{r,n}}xy=0
\end{equation}

\begin{equation}\label{c1.6}
{{g}_{1}}\left( x \right)=x-{{\tilde{T}}_{r,n}}\le 0,{{g}_{2}}\left( y \right)=y-{{\tilde{B}}_{r,n}}\le 0
\end{equation}

%
%

The Lagrange function is constructed by the Lagrange-multiplier method:


\begin{equation}\label{c1.8}
\mathcal{L}(x,y,\mathbf{\mu },\mathbf{\lambda })=x{{\Delta }_{\text{t}}}+y{{\Delta }_{\text{b}}}+{{\mu }_{1}}\left( {{N}_{r,n}}-{{a}_{r,n}}x-{{b}_{r,n}}xy \right)+{{\lambda }_{1}}\left( x-{{{\tilde{T}}}_{r,n}} \right)+{{\lambda }_{2}}\left( y-{{{\tilde{B}}}_{r,n}} \right) \\ 
\end{equation}
where KKT conditions are:

\begin{equation}\label{c1.9}
{{\mathcal{L}}_{x}}^{\prime }={{\Delta }_{\text{t}}}-{{a}_{r,n}}{{\mu }_{1}}-{{b}_{r,n}}{{\mu }_{1}}y+{{\lambda }_{1}}=0,{{\mathcal{L}}_{y}}^{\prime }={{\Delta }_{\text{b}}}-{{\mu }_{1}}{{b}_{r,n}}x+{{\lambda }_{2}}=0
\end{equation}

\begin{equation}\label{c1.10}
{{N}_{r,n}}-{{a}_{r,n}}x-{{b}_{r,n}}xy=0,{{\lambda }_{1}}\left( x-{{{\tilde{T}}}_{r,n}} \right)=0,{{\lambda }_{2}}\left( y-{{{\tilde{B}}}_{r,n}} \right)=0
\end{equation}

\begin{equation}\label{c1.11}
x-{{\tilde{T}}_{r,n}}\le 0,y-{{\tilde{B}}_{r,n}}\le 0,{{\lambda }_{1}}\ge 0,{{\lambda }_{2}}\ge 0
\end{equation}

%
%
%
%
%
%
%
%
%
%

\begin{enumerate}[a)]
\item When $x-{{\tilde{T}}_{r,n}}<0$ and $y-{{\tilde{B}}_{r,n}}<0$: ${{\lambda }_{1}}=0$, ${{\lambda }_{2}}=0$, union \ref{c1.9}, \ref{c1.10} to get: ${{x}_{1}}=\sqrt{\frac{{{N}_{r,n}}{{\Delta }_{\text{b}}}}{{{b}_{r,n}}{{\Delta }_{\text{t}}}}},{{y}_{2}}=\frac{1}{{{b}_{r,n}}}\left( \sqrt{\frac{{{N}_{r,n}}{{b}_{r,n}}{{\Delta }_{\text{t}}}}{{{\Delta }_{\text{b}}}}}-{{a}_{r,n}} \right)$. So when ${{N}_{r,n}}<N_{r,n}^{1}=\min \left\{ \frac{{{\left( {{{\tilde{T}}}_{r,n}} \right)}^{2}}b{{\Delta }_{\text{t}}}}{{{\Delta }_{\text{b}}}},\frac{{{\left( a+{{{\tilde{B}}}_{r,n}}b \right)}^{2}}{{\Delta }_{\text{b}}}}{b{{\Delta }_{\text{t}}}} \right\}$, $\left( {{x}_{1}},{{y}_{1}} \right)$ is the effective KKT point, otherwise the condition is not satisfied.
\end{enumerate}



\begin{enumerate}[b)]
\item When $x-{{\tilde{T}}_{r,n}}=0$ and $y-{{\tilde{B}}_{r,n}}=0$: ${{x}_{2}}={{\tilde{T}}_{r,n}},{{y}_{2}}={{\tilde{B}}_{r,n}}$. So when ${{N}_{r,n}}=N_{r,n}^{2}={{a}_{r,n}}{{\tilde{T}}_{r,n}}+{{b}_{r,n}}{{\tilde{T}}_{r,n}}{{\tilde{B}}_{r,n}}$, $\left( {{x}_{2}},{{y}_{2}} \right)$ is the effective KKT point, otherwise the condition is not satisfied.
\end{enumerate}


\begin{enumerate}[c)]
\item When $x-{{\tilde{T}}_{r,n}}<0$ and $y-{{\tilde{B}}_{r,n}}=0$: ${{x}_{3}}=\frac{{{N}_{r,n}}}{{{a}_{r,n}}+{{b}_{r,n}}{{{\tilde{B}}}_{r,n}}},{{y}_{3}}={{\tilde{B}}_{r,n}}$. So when ${{N}_{r,n}}<N_{r,n}^{3}={{\tilde{T}}_{r,n}}\left( {{a}_{r,n}}+{{b}_{r,n}}{{{\tilde{B}}}_{r,n}} \right)$, $\left( {{x}_{3}},{{y}_{3}} \right)$ is the effective KKT point, otherwise the condition is not satisfied.
\end{enumerate}


\begin{enumerate}[d)]
\item When $x-{{\tilde{T}}_{r,n}}=0$ and $y-{{\tilde{B}}_{r,n}}<0$: ${{x}_{4}}={{\tilde{T}}_{r,n}},{{y}_{4}}=\frac{{{N}_{r,n}}}{bx}-\frac{{{a}_{r,n}}}{{{b}_{r,n}}}$. So when ${{N}_{r,n}}<N_{r,n}^{4}={{b}_{r,n}}x\left( {{{\tilde{B}}}_{r,n}}+\frac{{{a}_{r,n}}}{{{b}_{r,n}}} \right)$, $\left( {{x}_{4}},{{y}_{4}} \right)$ is the effective KKT point, otherwise the condition is not satisfied.
\end{enumerate}


Based on constraints with different ${{N}_{r,n}}$, the qualified KKT points obtained from the above four cases are substituted into the original \ref{c1.4}, and the minimum point $\left( {{x}_{0}},{{y}_{0}} \right)$ of $f_{r,n}^{\text{d}\downarrow }\left( x,y \right)$ is the optimal resource scheduling strategy. $N_{r,n}^{2}=a{{\tilde{T}}_{r,n}}+b{{\tilde{T}}_{r,n}}{{\tilde{B}}_{r,n}}$ is the maximum size of data generation $N_{r,n}^{\max }\left( \mathbb{P}_{r,n}^{\text{d}\downarrow } \right)$ of $\mathbb{P}_{r,n}^{\text{d}\downarrow }$ in ${{\mathcal{T}}_{r}}$. If the value ${{N}_{r,n}}$ makes all four cases not KKT points, then the minimal resource scheduling strategy cannot be found, because the amount of available resources in the SRP can no longer meet the requirements. Therefore, the number of optimal resource scheduling strategies $\varpi _{r,n,r}^{*}\left( {{N}_{r,n}} \right)$ for data generation $\mathbb{P}_{r,n}^{\text{d}\downarrow }$ in ${{\mathcal{T}}_{r}}$ satisfies:

\begin{equation}\label{c1.22}
\left| \varpi _{r,n,r}^{*}\left( {{N}_{r,n}} \right) \right|=\left\{ \begin{aligned}
  & 1,\text{ }if\text{ }{{N}_{r,n}}\le N_{r,n}^{\max }\left( \mathbb{P}_{r,n}^{\text{d}\downarrow } \right) \\ 
 & 0,\text{ }if\text{ }{{N}_{r,n}}>N_{r,n}^{\max }\left( \mathbb{P}_{r,n}^{\text{d}\downarrow } \right) \\ 
\end{aligned} \right.
\end{equation}

\begin{enumerate}[2)]
\item For data consumption $\mathbb{P}_{r,n}^{\text{d}\uparrow }$ in ${{\mathcal{T}}_{r+1}}$:
\end{enumerate}

\begin{equation}\label{c1.23}
\begin{aligned}
  & \text{minimize: }f_{r,n}^{\text{d}\downarrow }\left( \mathbf{x},\mathbf{y},\mathbf{z} \right)={{\Delta }_{\text{t}}}\sum\limits_{{{x}_{i}}\in \mathbf{x}}{{{x}_{i}}}+{{\Delta }_{\text{b}}}\sum\limits_{{{y}_{i}}\in \mathbf{y}}{{{y}_{i}}}+{{\Delta }_{\text{f}}}\sum\limits_{{{z}_{i}}\in \mathbf{z}}{{{z}_{i}}} \\ 
 & =\left( {{x}_{1}}+{{x}_{2}}+{{x}_{3}} \right){{\Delta }_{\text{t}}}+\left( {{y}_{1}}+{{y}_{2}} \right){{\Delta }_{\text{b}}}+{{z}_{1}}{{\Delta }_{\text{f}}} \\ 
\end{aligned}
\end{equation}
S.T.

\begin{equation}\label{c1.24}
{{h}_{1}}\left( {{x}_{1}},{{y}_{1}} \right)=0,{{h}_{2}}\left( {{x}_{2}},{{z}_{1}} \right)=0,{{h}_{3}}\left( {{x}_{3}},{{y}_{2}} \right)=0
\end{equation}

\begin{equation}\label{c1.25}
{{g}_{1}}\left( {{x}_{1}},{{x}_{2}},{{x}_{3}} \right)\le 0,{{g}_{2}}\left( {{y}_{1}} \right)\le 0,{{g}_{3}}\left( {{y}_{2}} \right)\le 0,{{g}_{4}}\left( {{z}_{1}} \right)\le 0
\end{equation}

%
%
%
%
%
%
%
%
%
%
%
%

Construct the Lagrange function:

\begin{equation}\label{c1.31}
\mathcal{L}(\mathbf{x},\mathbf{y},\mathbf{z},\mathbf{\mu },\mathbf{\lambda })=\underbrace{f_{r,n}^{\text{d}\downarrow }\left( \mathbf{x},\mathbf{y},\mathbf{z} \right)}_{\text{Optimization function}}+\underbrace{\sum\limits_{{{h}_{i}}\left( \mathbf{x},\mathbf{y},\mathbf{z} \right)}{{{\mu }_{i}}{{h}_{i}}\left( \mathbf{x},\mathbf{y},\mathbf{z} \right)}}_{\text{Equality constraint}}+\underbrace{\sum\limits_{{{g}_{i}}\left( \mathbf{x},\mathbf{y},\mathbf{z} \right)}{{{\lambda }_{i}}{{g}_{i}}\left( \mathbf{x},\mathbf{y},\mathbf{z} \right)}}_{\text{Inequality constraint}}
\end{equation}

The determination of KKT conditions is similar to the specific proof process described above, and will not be repeated here, where the maximum size of data consumption $N_{r,n}^{\max }\left( \mathbb{P}_{r,n}^{\text{d}\uparrow } \right)$ can be obtained. As long as  ${{N}_{r,n}}\le N_{r,n}^{\max }\left( \mathbb{P}_{r,n}^{\text{d}\uparrow } \right)$, the optimal solution can be found. The number of optimal resource scheduling policies $\varpi _{r,n,r+1}^{*}\left( {{N}_{r,n}} \right)$ of data consumption $\mathbb{P}_{r,n}^{\text{d}\downarrow }$ in ${{\mathcal{T}}_{r+1}}$ meets:

\begin{equation}\label{c1.32}
\left| \varpi _{r,n,r+1}^{*}\left( {{N}_{r,n}} \right) \right|=\left\{ \begin{aligned}
  & 1,\text{ }if\text{ }{{N}_{r,n}}\le N_{r,n}^{\max }\left( \mathbb{P}_{r,n}^{\text{d}\uparrow } \right) \\ 
 & 0,\text{ }if\text{ }{{N}_{r,n}}>N_{r,n}^{\max }\left( \mathbb{P}_{r,n}^{\text{d}\uparrow } \right) \\ 
\end{aligned} \right.
\end{equation}

To sum up, the maximum transaction volume (MTV) for clients ${{u}_{n}}$ in IR ${{\mathcal{R}}_{r}}\in \mathcal{R}$ is defined as:

\begin{equation}\label{c1.33}
N_{r,n}^{\max }=\min \left\{ N_{r,n}^{\max }\left( \mathbb{P}_{r,n}^{\text{d}\downarrow } \right),N_{r,n}^{\max }\left( \mathbb{P}_{r,n}^{\text{d}\uparrow } \right) \right\}
\end{equation}

Therefore, with the limited resource constraint, SRP ${{\mathcal{M}}_{r,n}}=\left\{ \mathbf{b}_{r,n}^{\mathbf{t}},\mathbf{f}_{r,n}^{\mathbf{t}} \right\}$ and ${{\mathcal{M}}_{r+1,n}}=\left\{ \mathbf{b}_{r,n}^{\mathbf{t}},\mathbf{f}_{r,n}^{\mathbf{t}} \right\}$, when the MFP workload ${{N}_{r,n}}$ assumed by the client ${{u}_{n}}$ in ${{\mathcal{R}}_{r}}\in \mathcal{R}$ satisfied ${{N}_{r,n}}\in \left[ 0,N_{r,n}^{\max } \right]$, a resource scheduling strategy can always be found that minimizes the cost of local resources ${{c}_{r,n}}$.

\section{Graphical examples of MTV and MUTV.}
\label{C}

\begin{figure}[t]
 \centerline{\includegraphics[width=5in]{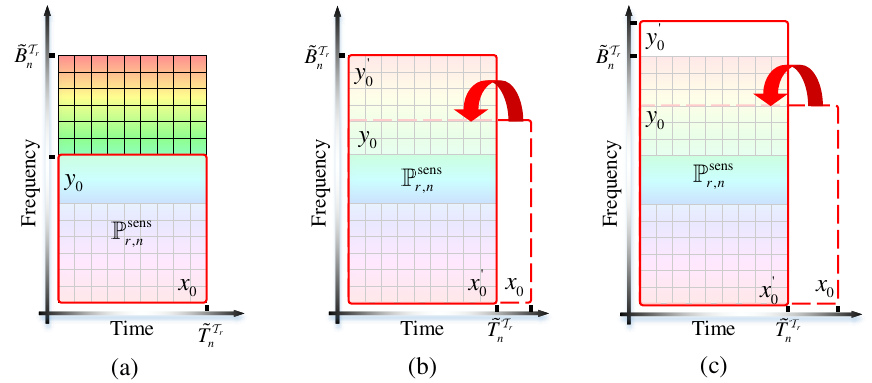}}
 \caption{Examples of complementarity between time-frequency resources.}
 \centering
 \label{fig14}
\end{figure}

\begin{enumerate}[1)]
\item If ${{x}_{0}}$ does not exceed the time-domain resource constraints, i.e. ${{x}_{0}}\le \tilde{T}_{n}^{{{\mathcal{T}}_{r}}}$, just execute the optimal local resource scheduling following $\left( {{x}_{0}},{{y}_{0}} \right)$, and the MUTV that meets this condition is $N_{r,n}^{\text{unc}}$, as shown in Fig. \ref{fig14}(a);
\item If ${{x}_{0}}$ exceeds the time-domain resource constraints, that is, ${{x}_{0}}>\tilde{T}_{n}^{{{\mathcal{T}}_{r}}}$, according to Appendix \ref{B}, the resource cost $c_{r,n}^{\text{d}\downarrow }$ of $\mathbb{P}_{r,n}^{\text{d}\downarrow }$ is still decreasing within the feasible domain $\left[ 0,\tilde{T}_{n}^{{{\mathcal{T}}_{r}}} \right]$ of time resources, that is:
\end{enumerate}

\begin{equation}\label{c1.34}
\left( c_{r,n}^{\text{d}\downarrow } \right)_{x}^{'}={{\Delta }_{\text{t}}}-\frac{{{N}_{r,n}}{{\Delta }_{\text{b}}}}{{{b}_{r,n}}}\frac{1}{{{x}^{2}}}<0
\end{equation}

So the cost function of the data generating $\mathbb{P}_{r,n}^{\text{d}\downarrow }$ is minimized at time ${{{x}'}_{0}}=\tilde{T}_{n}^{{{\mathcal{T}}_{r}}}$, when:

\begin{equation}\label{c1.35}
{{{y}'}_{0}}=\frac{{{N}_{r,n}}-{{a}_{r,n}}\tilde{T}_{n}^{{{\mathcal{T}}_{r}}}}{{{b}_{r,n}}\tilde{T}_{n}^{{{\mathcal{T}}_{r}}}}
\end{equation}

\begin{enumerate}[a)]
\item If $0\le {{{y}'}_{0}}\le \tilde{B}_{n}^{{{\mathcal{T}}_{r}}}$, the solution $\left( {{{{x}'}}_{0}},{{{{y}'}}_{0}} \right)$ is feasible, i.e. the available resources of SRP in a can complete the data generation volume of ${{N}_{r,n}}$, as shown in Fig. \ref{fig14}(b); Specifically, define ${{N}_{r,n}}$ that satisfies ${{{y}'}_{0}}=\tilde{B}_{n}^{{{\mathcal{T}}_{r}}}$ as the MTC, which is:
\end{enumerate}

\begin{equation}\label{c1.36}
N_{r,n}^{\max }\left( \mathbb{P}_{r,n}^{\text{d}\downarrow } \right)\triangleq {{N}_{r,n}}|{{x}_{0}}=\tilde{T}_{n}^{{{\mathcal{T}}_{r}}},{{y}_{0}}=\tilde{B}_{n}^{{{\mathcal{T}}_{r}}}
\end{equation}

\begin{enumerate}[b)]
\item If ${{{y}'}_{0}}>\tilde{B}_{n}^{{{\mathcal{T}}_{r}}}$, SRP resources of client ${{u}_{n}}$ in ${{\mathcal{T}}_{r}}$ cannot meet the requirements of the data generation $\mathbb{P}_{r,n}^{\text{d}\downarrow }$ where workload is ${{N}_{r,n}}$, as shown in Fig. \ref{fig14}(c).
\end{enumerate}

The data consumption $\mathbb{P}_{r,n}^{\text{d}\uparrow }$ in ${{\mathcal{T}}_{r+1}}$ similar, and the maximum transaction direction for a client in the MFP service market at IR is $N_{r,n}^{\max }\left( \mathbb{P}_{r,n}^{\text{d}\uparrow } \right)$:

To sum up:

\begin{equation}\label{c1.37}
N_{r,n}^{\max }=\min \left\{ N_{r,n}^{\max }\left( \mathbb{P}_{r,n}^{\text{d}\downarrow } \right),N_{r,n}^{\max }\left( \mathbb{P}_{r,n}^{\text{d}\uparrow } \right) \right\}
\end{equation}

\end{document}